\documentclass[pre,twocolumn,graphicx]{revtex4-1}

\usepackage{amsmath}
\usepackage{mathtools}
\usepackage{amssymb}
\usepackage{verbatim}   
\usepackage{color}      
\usepackage{subfigure}  
\usepackage{hyperref}   
\usepackage{cancel}
\usepackage{soul,xcolor}

\newcommand{\bnabla}{\mbox{\boldmath $\nabla$}}

\newcommand{\ba}{\begin{eqnarray}}
\newcommand{\ea}{\end{eqnarray}}
\newcommand{\be}{\begin{equation}}
\newcommand{\ee}{\end{equation}}

\setstcolor{red}

\begin{document}

\title{Mean-field theory of active electrolytes:  Dynamic adsorption and overscreening}

\author{Derek Frydel}
\affiliation{Department of Chemistry, Federico Santa Maria Technical University, Campus San Joaqu\'in, Santiago, Chile}
\author{Rudolf Podgornik}
\affiliation{School of Physical Sciences and Kavli Institute for Theoretical Sciences,
University of Chinese Academy of Sciences, Beijing 100049, China\\
CAS Key Laboratory of Soft Matter Physics, Institute of Physics,
Chinese Academy of Sciences, Beijing 100190, China\\
Department of Physics, Faculty of Mathematics and Physics, University of Ljubljana, 1000 Ljubljana, Slovenia\\ 
Department of Theoretical Physics, J. Stefan Institute, 1000 Ljubljana, 1000 Ljubljana, Slovenia}

\date{\today}

\begin{abstract}
We investigate active electrolytes within the mean-field level of description. The focus is on how the 
double-layer structure of passive, thermalized charges is affected by active dynamics of constituting 
ions. One feature of active dynamics is that particles adhere to hard surfaces, regardless of chemical 
properties of a surface and specifically in complete absence of any chemisorption or physisorption. 
To carry out the mean-field analysis of the system that is out of equilibrium, we develop the 
"mean-field simulation" technique, where the simulated system consists of charged parallel sheets 
moving on a line and obeying active dynamics, with the interaction strength rescaled by the number of 
sheets. The mean-field limit becomes exact in the limit of an infinite number of movable sheets.
\end{abstract}
\pacs{
}

\maketitle

\section{Introduction}

The mean-field approximation is the most common tool in dealing with electrostatic systems in the 
weak-coupling regime \cite{Rudi13,David18,Yan02}.  The suitability of the mean-field collective description 
to electrostatics stems from the long-range nature of Coulomb interactions;  because each particle interacts 
with all other particles in a system; this, in turn, brings about significant suppression of fluctuations in 
systems under standard conditions \cite{Frydel16c,Frydel17}.

In recent years, the mean-field analysis has been extended to ions with inner structure to include 
an ever larger class of emerging systems.  These extensions incorporate steric effects 
\cite{David97,Frydel11a,Frydel12,Frydel12,Frydel14}, multipolar interactions \cite{David07,Rudi09}, 
polarizability \cite{Frydel11,Hatlo12}, penetrability 
\cite{Hansen11a,Hansen11b,Frydel13,Masters13,Levesque14,Frydel16a,Frydel18}, non-spherical 
shapes such as that of dumbbell 
ions \cite{Bohinc04,Bohinc08,Bohinc11,Bohinc12}, and the various combinations of the above extensions 
\cite{Frydel16b}.

\begin{figure}[h] 
 \begin{center}
 \begin{tabular}{rr}
\hspace{-1cm}  \includegraphics[height=0.18\textwidth,width=0.23\textwidth]{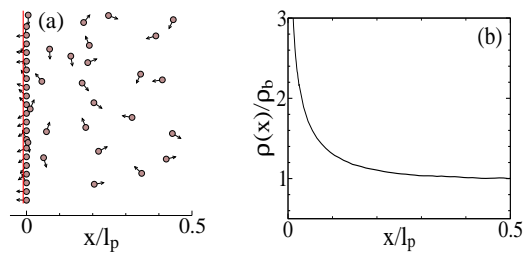}&
  \hspace{0.5cm}\includegraphics[height=0.18\textwidth,width=0.20\textwidth]{rho_wall.eps}\\
 \end{tabular}
 \end{center}
\caption{A snapshot of active ideal particles in two dimensions near a confining hard wall at $x=0$, 
and (b) the corresponding stationary distribution normalized by a bulk density $\rho_b$ . In overdamped dynamics 
particles are not reflected in contact with a wall; instead they remain trapped until an orientational vector 
diffuses to the direction away from a wall. Released particles have initially almost parallel orientation with 
a wall. This causes a released particle to "linger" in the vicinity of a wall, giving rise to a divergent density profile.}
\label{fig:pic}
\end{figure}

In the present work we extend the mean-field approximation not to different ionic structures, but to different 
dynamics. Thus, point ions instead of being Brownian are now active particles, with trajectories characterized 
by constant velocity and diffusing orientation, giving rise to orientational persistence. 
A peculiar feature of active dynamics, and the direct consequence of orientational persistence combined with 
overdamped dynamics, is that particles adhere to hard surfaces \cite{Lee13} (see Fig. (\ref{fig:pic})).  
Such "dynamic adsorption" is different from chemisorption and physisorption, which depend on the chemistry of 
a surface  
\cite{Ninham71,White75,Rudi14,Rudi15,Tomer16,Tomer18,Blum89,McQuarrie95,McQuarrie96a,McQuarrie96b}.  
\graphicspath{{figures/}}

The signature of "dynamic adsorption" is the emergence of a divergent density profile at the location of a 
hard surface. From the mathematical point of view, the divergence ensures the continuity of a density, at 
the point where the density profile of free particles merges with the Dirac delta function representing the 
distribution of adsorbed particles. Physically, the divergence arises due to newly released particles, 
which tend to have orientations nearly parallel to a wall.  Such particles have a tendency to "linger" 
in the vicinity of a wall, giving rise to an observed divergence.

In case of ions, "dynamic adsorption" will naturally modify the surface charge of a wall. It is the aim of 
this paper to investigate this phenomenon as well as other dynamic contributions to the structure and 
properties of a double-layer, within the mean-field level of description.

Part of the challenge in achieving these goals is technical. The stationary distributions of 
active particles cannot 63 be obtained from their corresponding Boltzmann weights. 
In consequence, even the simple case of noninteracting active particles is not amenable to analytical 
solutions based on the 66 Fokker-Planck equation \cite{Fisch80,Cates15c,Wagner17}, 
and the dynamic simulations still offer the most direct way to analyze these systems.  

In the present work we develop the "mean-field simulation," which allows us to study charged systems 
with plane geometry within the mean-field level of description. The simulated system consists of charged 
sheets moving on a line;  thus, it is strictly one-dimensional. The mean-field limit corresponds to infinitely 
many sheets with an infinitesimal surface charge, while the total charge density remains fixed. The method 
is designed specifically for planar geometry. The "mean-field simulation" developed in this work is, therefore,
not directly transferable to other geometries.

The present work is organized as follows. In Sec. \ref{sec:preliminaries} we review general properties of 
active particles. In Sec. \ref{sec:gravity} we go over some of the results for active particles in a gravitational field. 
In Secs. \ref{sec:ocp} and Sec. \ref{sec:electrolyte} we develop the mean-field simulation for active ions. 
In Secs. \ref{sec:ocp} and \ref{sec:electrolyte} we proceed to apply that method to study various systems and 
their variations. Finally, in Sec. \ref{sec:conclusion} we conclude the work.


\section{Preliminaries}
\label{sec:preliminaries}

The motion of an active-Brownian particle is characterized by a constant velocity $v_0$ and a diffusion 
of an orientational vector ${\bf n}(\theta,\phi)=(\sin\theta\cos\phi,\sin\theta\sin\phi,\cos\theta)$ over a spherical
surface of unit radius with a rotational diffusion constant $D_r$ (units of 1/time). Diffusion of an orientation 
vector leads to directional persistence, resulting in the exponentially decaying orientation-orientation correlation 
function $\langle{\bf n}(0)\cdot{\bf n}(t)\rangle=e^{-t/\tau_p}$, where

\be
\tau_p = \frac{1}{d-1}\frac{1}{D_r}, 
\ee
is the persistence time and $d$ is a system dimensionality \cite{Lowen16}.  
This leads to a translational persistence with the persistence length given by 
\be
l_p = v_0\tau_p. 
\ee
If a distance that an active particle covers in time $t$ is 
$
{\bf r}(t) = v_0\int_0^tds\,{\bf n}(s),   
$
then the expression for the mean-squared displacement evaluates as 
\ba
\langle r^2(t)\rangle 
&=& 2l_p^2\bigg[\frac{t}{\tau_p} + e^{-t/\tau_p}-1\bigg].
\label{eq:msd}
\ea
The above expressions spans two limits.  At short times the motion is ballistic, 
$\langle r^2(t)\rangle \approx l_p^2(t/\tau_p)^2$.  Then at long times it is diffusive, 
\be
\langle r^2(t)\rangle \approx \bigg(\frac{l_p^2}{\tau_p}\bigg) t, 
\ee
with the "effective" translational diffusion constant given by 
\be
D_{\rm eff} = \frac{1}{d}\frac{l_p^2}{\tau_p}.  
\label{eq:Deff}
\ee
Using the Einstein relation $D=k_BT/\zeta$, where $\zeta$ designates the friction coefficient 
of a medium, it becomes possible to define an "effective" temperature:  
\be
k_BT_{\rm eff} = \frac{1}{d}\frac{\zeta l_p^2}{\tau_p}. 
\label{eq:kT}
\ee
The above temperature expression will be used in subsequent sections for comparing the 
results for active particles with those for passive particles.

The temperature expression in Eq. (\ref{eq:kT}) is not to be confused with the equilibrium temperature of 
a medium in which active particles are immersed.  The medium contributes two effects.  First, it is perfectly 
dissipative, leading to overdamped (non-Newtonian) dynamics.  Second, it contributes thermal fluctuations.  
In the present study, the temperature of a medium is set to zero, in order to isolate the contributions due to 
active dynamics. 

To emphasize the essential role of overdamped dynamics in the phenomena of dynamic adsorption, we 
consider a particle propagating through a scattering medium, that is, a medium of randomly distributed, 
small-angle, elastic point scatterers  \cite{Fisch80}. The velocity of a scattered particle is constant while its 
direction fluctuates. Up to this point, this is precisely what occurs for active particles. The scattering medium, 
however, is not dissipative, and dynamically the system is Newtonian.  The two systems start to deviate in 
presence of external potentials or interparticle interactions. In consequence, the scattering system does not 
produce dynamic adsorption.

\subsection{Harmonic confinement}
To demonstrate the significance of the persistence length, we consider a single active particle inside 
a harmonic trap \cite{Brady16}.  
We define the size of a trap, $l_k$, as a distance measured from the trap center, beyond which an active particle cannot 
penetrate.  This distance corresponds to the balance between the two forces:  
the particle intrinsic force, $\zeta v_0$, and the trap's 
restoring force, $F=-kr$.  The condition of balance yields $l_k=\zeta v_0/k$.

\graphicspath{{figures/}}
\begin{figure}[h]
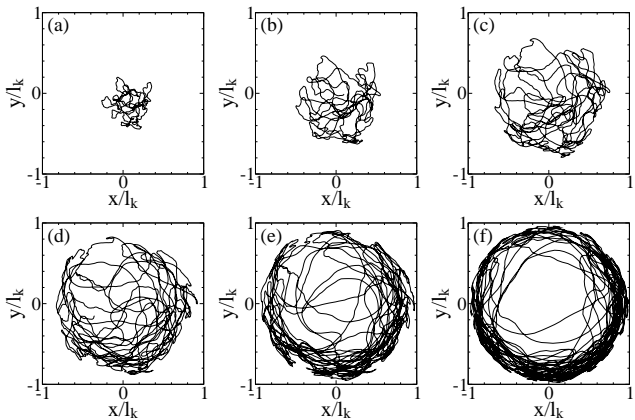
 
 \begin{center}
 \begin{tabular}{llllll}
  \includegraphics[height=0.15\textwidth,width=0.15\textwidth,clip=true]{c_k01Ca.eps} &
  \includegraphics[height=0.15\textwidth,width=0.15\textwidth,clip=true]{c_k025Ca.eps} &
  \includegraphics[height=0.15\textwidth,width=0.15\textwidth,clip=true]{c_k05Ca.eps} \\
    \includegraphics[height=0.15\textwidth,width=0.15\textwidth,clip=true]{c_k1Ca.eps} &
  \includegraphics[height=0.15\textwidth,width=0.15\textwidth,clip=true]{c_k2Ca.eps} &
  \includegraphics[height=0.15\textwidth,width=0.15\textwidth,clip=true]{c_k4Ca.eps} \\
 \end{tabular}
 \end{center}
\caption{Trajectories for a single active particle in a harmonic trap for different ratios $l_p/l_k$
($0.1,0.25,0.5,1,2,4$).  The simulation is carried out in 2D to facilitate visualization.  
The duration of each trajectory is $100\tau_p$.  }
\label{fig:c}
\end{figure}
Fig. (\ref{fig:c}) shows a number of trajectories of an active particle inside a trap for different ratios $l_p/l_k$.
The figures are plotted in units of $l_k$.  As confinement increases, the trajectories become increasingly 
restricted to the border region of a trap at $r/l_k=1$.  Fig. (\ref{fig:rho_c}) shows the corresponding stationary 
distributions.  
\graphicspath{{figures/}}
\begin{figure}[h]
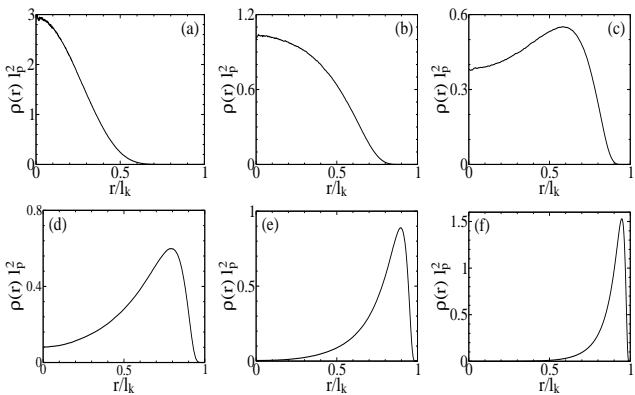
 
 \begin{center}
 \begin{tabular}{rrrrrr}
  \includegraphics[height=0.14\textwidth,width=0.15\textwidth]{rho_k01B.eps} &
  \includegraphics[height=0.14\textwidth,width=0.15\textwidth]{rho_k025B.eps} &
  \includegraphics[height=0.14\textwidth,width=0.15\textwidth]{rho_k05B.eps} \\
  \includegraphics[height=0.14\textwidth,width=0.15\textwidth]{rho_k1B.eps} &
  \includegraphics[height=0.14\textwidth,width=0.15\textwidth]{rho_k2B.eps} &
  \includegraphics[height=0.135\textwidth,width=0.15\textwidth]{rho_k4B.eps} \\
 \end{tabular}
 \end{center}
\caption{Stationary distributions corresponding to Fig. (\ref{fig:c}). }
\label{fig:rho_c}
\end{figure}
For the lowest ratio $l_p/l_k$, the distribution is nearly Gaussian, $\sim e^{-kr^2/(2k_BT)}$, 
with $k_BT$ given in Eq. (\ref{eq:kT}).  As the ratio $l_p/l_k$ decreases, the distribution shifts 
toward the edge of a wall.  For $l_p/l_k=4$, an active particle is almost entirely confined to the 
edge of a trap.

\begin{figure}[h]
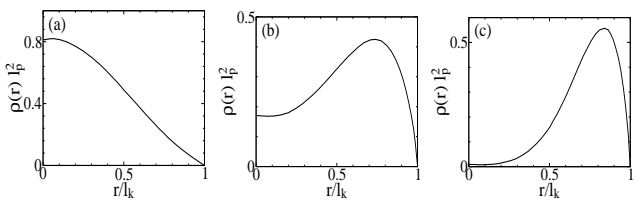
 
 \begin{center}
 \begin{tabular}{rrrrrr}
  \includegraphics[height=0.14\textwidth,width=0.15\textwidth]{rho_worm_k2.eps} &
  \includegraphics[height=0.14\textwidth,width=0.15\textwidth]{rho_worm_k4.eps} &
  \includegraphics[height=0.135\textwidth,width=0.15\textwidth]{rho_worm_k8.eps} \\
 \end{tabular}
 \end{center}
\caption{Distribution of a wormlike chain inside a circular hard-wall trap with radius $R=l_k$.  Figures are 
from Ref. \cite{Chen2014} for $l_p/l_k=2,4,8$.  As $l_p/l_k\to 0$ (the weak confinement limit) the distributions 
approach those of a Gaussian chain model.   }
\label{fig:rho_worm}
\end{figure}

The situation is analogous to the packing problem of a wormlike chain inside a circular hard-wall confinement 
\cite{Chen2014,Chen2016}, see Fig. (\ref{fig:rho_worm}).  As the persistence length of a wormlike chain 
increases, a polymer becomes less elastic, and the energetically least costly configuration corresponds to 
packing along the inner circumference of a trap.  
\graphicspath{{figures/}}

A similar breakdown of Boltzmann statistics is expected to occur in any system with some 
confinement length, at the point where the confinement length becomes comparable with the persistence
length.

\subsection{Dynamic simulation} 
\subsubsection{Dynamic integration}

Dynamic simulations use the time integration, for $z(t)$ and $\theta(t)$, updated at discrete time steps 
every $\Delta t$.  The present work uses the Euler integrator, which is the simplest update algorithm, 
\be
z(t+\Delta t) = z(t) + \bigg[ v_0 \cos\theta(t) + \frac{F(z(t))}{\zeta}\bigg]  \Delta t, 
\label{eq:z}
\ee
\be
\theta(t+\Delta t) = \theta(t) + \hat\xi(t)\sqrt{2D_r \Delta t} + D_r \cot\theta(t)\Delta t.
\label{eq:theta}
\ee
$F(z)$ in the first equation denotes an external force.  
Then the term $\hat\xi(t)$ in the second equation represents a random values generated at each time step 
and taken from a Gaussian distribution with zero and unity variance.  
The second equation contains a deterministic term $D_r\cot\theta$, even if there is no such 
deterministic torque \cite{Hsu03,Engel04,Gompper15}.  This term is not present in $d=2$ and it arises 
in $d=3$ to account for a spherical surface over which a vector ${\bf n}$ diffuses.  The integration
over $x(t)$, $y(t)$, and $\phi(t)$ are not required on account of a wall geometry.  

Straightforward integration in Eq. (\ref{eq:theta}) can be problematic for orientations near the two 
poles, $\theta=0$ and $\theta=\pi$, where the deterministic term diverges.  This problem is corrected 
by integrating instead an orientation vector in Cartesian coordinates.  Since for the wall geometry 
only $n_z$ is relevant, the Euler integrator for this component of an orientation is \cite{Gompper15}, 
\ba
n_z(t+\Delta t) &=& n_z(t) - \hat\xi_1(t)\sin\theta(t) \sqrt{2D_r\Delta t} \nonumber\\ 
&-& D_r\big(\hat\xi_{1}^2(t) + \hat\xi_{2}^2(t)\big)n_z(t)\Delta t,
\label{eq:nz}
\ea
where $n_z=\cos\theta$.  Note that two random numbers $\hat\xi_1(t)$ and $\hat\xi_2(t)$ are generated at each time step.

\subsubsection{boundary conditions}
\label{sec:BC}

Dynamic simulations are carried out within an interval $z \in(0,L)$.  Each time a particle crosses a 
boundary at $z=0$ or $z=L$, it is moved to the exact location of a boundary in the same step.  
A possibility of a particle being bounced from a wall is prevented by overdamped dynamics.

Because a particle can move away from a wall only if its orientation points in the direction away from a wall, 
and because such reorientation does not occur instantaneously due to orientational relaxation, 
a particle that arrives at a wall is considered as adsorbed, at least for the time being. 
This procedure is repeated until a velocity vector points away from a wall.

The interval length of a simulation box, $L$, depends on a particular situation.  For a gravitational 
system considered in the next section $L\to\infty$, as particles are naturally confined by a gravitational force.

\subsubsection{reduced units}
For the unit of length and time we take $l_p$ and $\tau_p$.  Consequently, the reduced position 
on the $z$-axis is $z^*=z/l_p$, the reduced time is $t^*=t/\tau_p$, and a reduced density is $\rho^*=l_p^3\rho$.  
Furthermore, the unit of force is $\zeta v_0$ and the reduced force is $F^*=F/(\zeta v_0)$ and
the reduced pressure is $P^*=P l_p^2/(\zeta v_0)$.
The time step that we use is $\Delta t^*=0.001$.

\section{gravitational force}
\label{sec:gravity}

Prior to considering charged particles, we review results for non-interacting active particles 
in the gravitational field, $F=-G$.  The Euler integrators for this situation for the position and 
orientation vector are 
\be
z^*(t^*+\Delta t^*) = z^*(t^*) + \big[n_z(t^*) -   \gamma_G\big]\Delta t^*,
\label{eq:zg}
\ee
where the reduced gravitational force is 
\be
\gamma_G = \frac{G}{\zeta v_0}, 
\label{eq:gammaG}
\ee   
and 
\ba
n_z(t^*+\Delta t^*) &=& n_z(t) - \hat\xi_1(t)\sin\theta(t^*) \sqrt{\Delta t^*} \nonumber\\ 
&-& \frac{1}{2}\big(\hat\xi_{1}^2(t^*) + \hat\xi_{2}^2(t^*)\big)n_z(t^*)\Delta t^*,
\ea
both equations expressed in reduced units.  

Ideal active particles in gravitational field in 2D have been considered in Ref. \cite{Cates15c,Wagner17}.  
The analysis was based on separation of variables, leading to a Mathieu differential equation 
for the orientational counterpart \cite{Mathieu1868,Ruby96,Frenkel01}.  A dominant exponential decay could 
be obtained from the lowest eigenfunction.  
For active particles in gravitational field in 3D space, separation of variables does not generate the Mathieu 
equation, and it is not clear if analytical results in terms of known functions are available.  
Here we don't attempt analytical analysis and limit ourselves to a simple presentation of simulation results.

\begin{figure}[h] 
 \begin{center}
 \begin{tabular}{rrrr}
  \includegraphics[height=0.19\textwidth,width=0.23\textwidth]{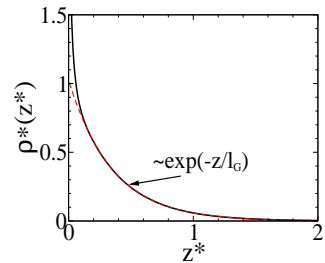}&
 \end{tabular}
 \end{center}
\caption{The stationary distribution of active-Brownian particles in the 
gravitational field with strength $\gamma_G=0.5$.  The system is confined by a wall 
at $z^*=0$. The profile is obtained from a dynamic simulation and the fit at a far-field 
demonstrates exponential decay.  }
\label{fig:rho} 
\end{figure}

Fig. (\ref{fig:rho}) shows an average stationary distribution for the gravitational strength $\gamma_G=0.5$.  
The profile exhibits two distinct regions, the near-field divergence (implying dynamic adsorption) 
followed by an exponential decay.  
Due to adsorption, the integrated density of free particles is less than the  
total number of particles in a system, $\int_0^{\infty}dz^*\,\rho^*(z^*)<n$.  
\graphicspath{{figures/}}

\begin{figure}[h] 
 \begin{center}
 \begin{tabular}{rrrr}
  \includegraphics[height=0.19\textwidth,width=0.23\textwidth]{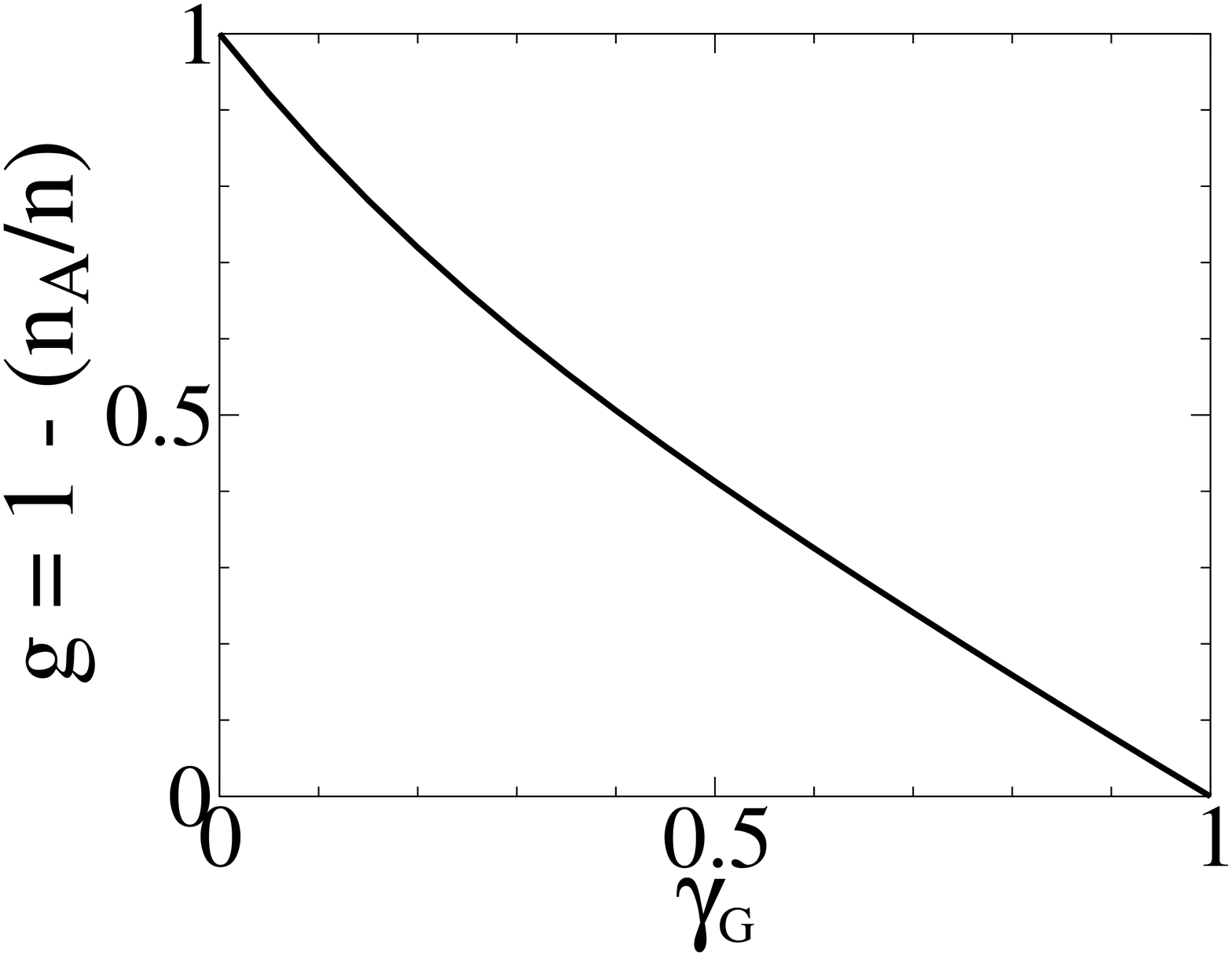}&
 \end{tabular}
 \end{center}
\caption{The fraction of free (non-adsorbed) particles as a function of $\gamma_G$. } 
\label{fig:NA} 
\end{figure}

In Fig. (\ref{fig:NA}) we plot the quantity $g=1-n_A/n$, $g$ corresponds to a fraction of free particles, 
as a function of $\gamma_G$.  
\graphicspath{{figures/}}
For no adsorbed particles $g=1$, and for all adsorbed particles $g=0$.   
The monotonically decreasing behavior of $g$ is expected, but the surprising feature of the plot 
is that $g$ vanishes beyond $\gamma_G=1$.  For $\gamma_G=1$ and beyond all the particles 
are adsorbed.  The adsorption is complete and permanent.  This has a very straightforward explanation.  
For $\gamma_G>1$ particle velocities in the $z$-direction can only be negative, regardless of their 
orientation.  But because particles cannot go beyond $z=0$, they become permanently trapped.  
Their motion occurs only in the plane of a wall.

Next we consider the far-field behavior of a distribution.  The fact that it is exponential is
not particularly surprising, that is, $\rho(z)\sim e^{-z/l_G}$.  Of more interest is how the scaling 
of the decay length $l_G$ scales with $\gamma_G$.  It is reasonable to assume that in the limit 
$\gamma_G\to 0$ one recovers the Boltzmann-like behavior, where $\rho(z)\sim e^{-G z/k_BT}$,
because $l_G$ and therefore the confinement is large, $l_G\to\infty$.  
Substituting 
for $k_BT$ in Eq. (\ref{eq:kT}) we get $\rho^*(z^*) \sim e^{- 3\gamma_G z^*}$, so that the decay 
length is 
\be
\lim_{\gamma_G\to 0}\frac{l_G}{l_p} = \frac{1}{3\gamma_G}.  
\label{eq:lambda_G}
\ee
This scaling is confirmed by simulations for small $\gamma_G$.  But as $\gamma_G$ increases, 
and the confinement decreases, 
$l_G$ start to deviates from the behavior in Eq. (\ref{eq:lambda_G}).  To capture the deviations
from the Boltzmann behavior, we introduce the dimensionless parameter $g_G$, referred to as the 
renormalization constant, and rewrite the expression in Eq. (\ref{eq:lambda_G}) as 
\be
\frac{l_G}{l_p} = \frac{g_G}{3\gamma_G}.  
\label{eq:gG}
\ee

\begin{figure}[h] 
 \begin{center}
 \begin{tabular}{rrrr}
  \includegraphics[height=0.19\textwidth,width=0.23\textwidth]{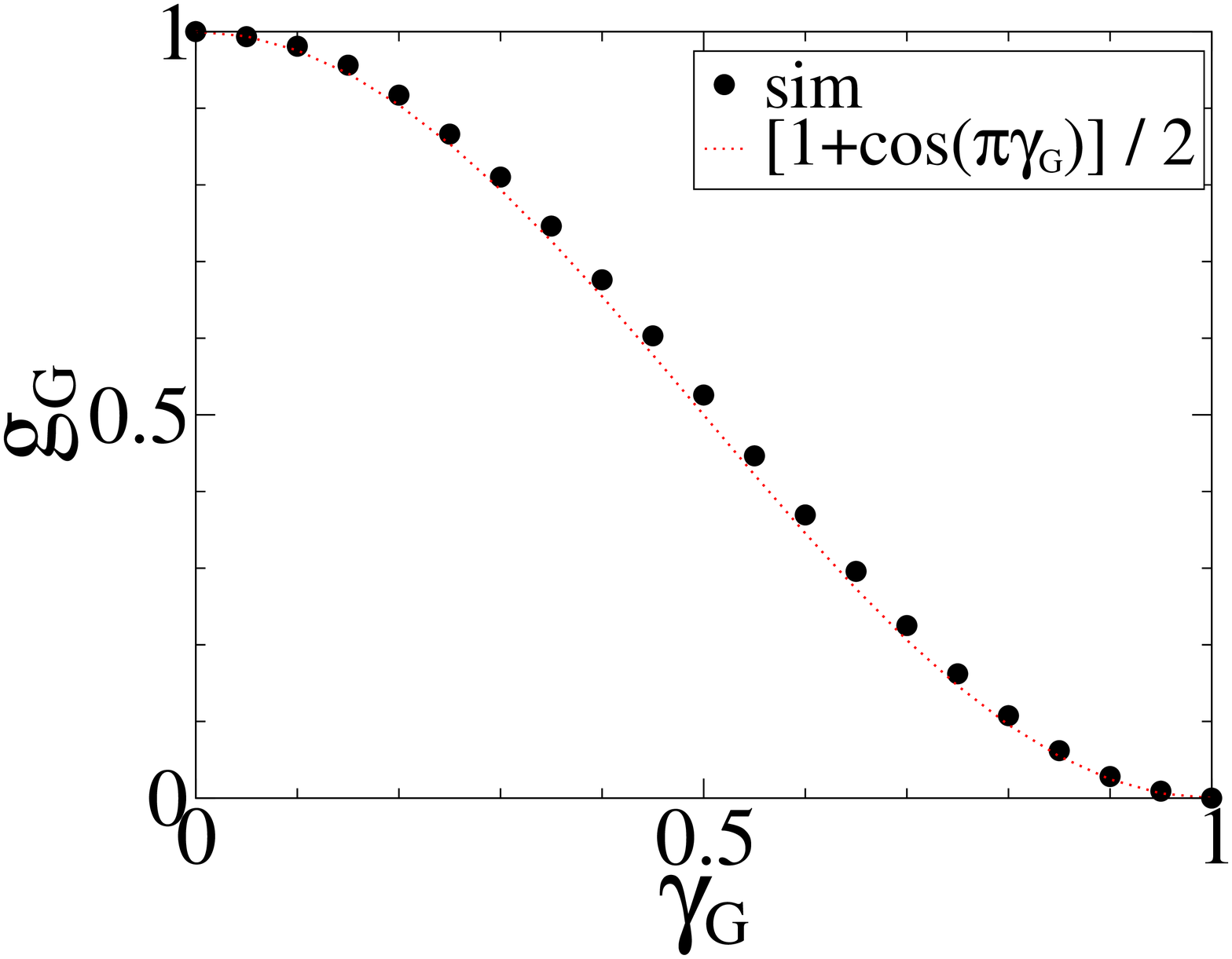}&
 \end{tabular}
 \end{center}
\caption{The renormalization constant $g_G$, introduced in Eq. (\ref{eq:gG}), as a function of  
$\gamma_G$.  The data points are compared to the empirical functional form 
$(1+\cos\gamma_G\pi)/2$.  } 
\label{fig:gG} 
\end{figure}

In Fig. (\ref{fig:gG}) we plot $g_G$ as a function of $\gamma_G$.  
The observation is that $g_G$ decreases with the strength of a gravitational force.  The
explanation is that as $\gamma_G$ increases, the 
orientations that generate positive velocity (velocity away from a wall) become reduced.  
\graphicspath{{figures/}}
The behavior of $g_G$ is roughly trigonometric, $(1+\cos\gamma_G\pi)/2$.  Beyond $\gamma_G= 1$, $g_G$ 
vanishes.

As the final point, we investigate the link between adsorbed particles and pressure 
\cite{Brady14,Cates15a,Cates15b,Dean18}.  In the case of active particles, pressure is calculated 
by counting the number of adsorbed particles and then taking into account their orientation.  
The reason is that only adsorbed particles exert force on a wall.  The force is 
proportional to a particle velocity in the direction toward a wall.  In reduced units, this is expressed as 
\be
\frac{l_p^2 P}{\zeta v_0} = 2\pi n_A \int_{0}^{\pi}  d\theta\, \sin\theta \big(\gamma_G - \cos\theta) \, \rho_A(\theta),
\ee 
where $\rho_A(\theta)$ is the angular distribution of adsorbed particles, normalized as 
$2\pi \int_0^{\pi}d\theta\, \sin\theta\rho_A(\theta)=1$.  One could evaluate the integral by considering the 
Fokker-Planck equation of the present system, see Appendix (\ref{sec:anal0}) for details.  However, 
knowing that a force exerted by active particles must be equal and opposite to the gravitational force exerted on 
the same particles, we guess the following result
\be
\frac{l_p^2 P}{\zeta v_0} = \gamma_G n, 
\ee
which is confirmed by simulations.

\section{The mean-field treatment of active ions}
\label{sec:mf}

In the case of active ions, there is no gravitational field, but a similar constant force arises from a uniform 
surface charge, $\sigma_c$, interacting with ions on account of their charge $\pm q$.  In addition to the one-body
constant force, there are now many-body interactions due to Coulomb forces between ions.

Within the mean-field approximation, an ion, instead of interacting with individual ions, interacts with an 
average charge distribution due to all ions.  This reduces a many-body to a one-body problem.  
Given a two-component symmetric electrolyte $q:q$, the averaged one-body force is 
\be
F_{\pm}(z) = \mp q_{} \frac{d\psi(z)}{dz},
\ee
where $\psi(z)$ is the average electrostatic potential. Using the Poisson relation,  
\be
\epsilon \frac{d^2\psi(z)}{dz^2} = -\rho_c(z),
\ee
where $\rho_c(z) = q\rho_+(z) - q\rho_-(z)$, 
$\rho_{\pm}(z) = \int_0^{\pi}d\theta\,\rho_{\pm}(z,\theta)$, and $\epsilon$ is the dielectric constant
of a medium, the force can be expressed in terms of a charge density distribution as 
\be
F_{\pm}(z) = \mp\frac{q}{\epsilon}\int_{z}^{\infty} dz'\,\rho_c(z'), 
\label{eq:f}
\ee
where the neutrality constraint requires 
\be
\int_0^{\infty} dz\,\rho_c(z) = -\sigma_c.  
\ee

\section{The mean-field simulation}
\label{sec:mf_sim}

For the case of passive ions, the mean-field framework is completed by expressing the charge 
density using the Boltzmann weight, $\rho_c\approx qc_s (e^{-\beta q\psi}-e^{\beta q\psi})$, leading
to the Poisson-Boltzmann equation.  For active ions there is no equivalent of the Boltzmann factor, 
and there is no available expression for $\rho_c$ in terms of $\psi$.  

To overcome this problem, we work with instantaneous distributions, 
\be
\hat\rho_{+}(z) = \bigg(\frac{\sigma_c}{qn}\bigg) \sum_{i=1}^{n_s}\delta(z_i^{+}-z), 
\label{eq:rho+}
\ee
\be
\hat\rho_{-}(z) = \bigg(\frac{\sigma_c}{qn}\bigg) \sum_{i=1}^{n+n_s}\delta(z_i^{-}-z),
\label{eq:rho-}
\ee
where the weight factor $\sigma_c/(qn)$ determines the contribution of each particle to a density
and assures correct units.  Note that there are more counterions than coions, $n_-=n_s+n$ and 
$n_+=n_s$, where $n_s$ is the number of salt ions and $n$ is the excess of counterions
needed to neutralize the surface charge $\sigma_c\ge 0$.  The mean-field distributions correspond 
to averaged instantaneous distributions,  
\be
\rho_{\pm}^{\rm mf}(z) = \Big\langle \hat\rho_{\pm}(z) \Big\rangle. 
\ee

Using the above expressions, the instantaneous charge density, defined as 
$\hat\rho_c(z)=q\hat\rho_+(z) - q\hat\rho_-(z)$, is written as 
\be
\hat\rho_c(z) = q \bigg(\frac{\sigma_c}{qn}\bigg) \bigg[   \sum_{i=1}^{n_s}\delta(z_i^{+}-z) - \sum_{i=1}^{n_s+n}\delta(z_i^{-}-z)    \bigg], 
\ee
where the charge of each particle is rescaled by $n$ and the total charge density correctly integrates as
\be
\int_0^{\infty} dz\,\hat\rho_c(z) = -\sigma_c,
\ee
satisfying the neutrality constraint.

Now it becomes possible to write the expression of force, analogous to that in Eq. (\ref{eq:f}), 
\be
F(z_i^{\pm}) =  \mp\frac{q}{\epsilon} \frac{\sigma_c}{n} A(z_i^{\pm}), 
\label{eq:F}
\ee
where 
\be
A(z_i^{\pm}) = \sum_{j=1}^{n_s} H(z_i^{\pm}-z_j^+) - \sum_{j=1}^{n_s+n} H(z_i^{\pm}-z_j^-),
\label{eq:A}
\ee
and
\be
H(z) = \left\{ 
  \begin{array}{ccc}
     1, & \quad \text{if $z>0$}\\
     \frac{1}{2}, & \quad \text{if $z= 0$}\\
     0, & \quad \text{if $z< 0$}
        \label{eq:H}
  \end{array} 
  \right.
  \ee
is the step function.  The term $\sum_{j} H(z_i-z_j)$ counts the number of particles within the interval 
$z_i\le z<\infty$.  Because $H(0)=1/2$, a particle that defines the lower limit of the interval 
contributes one half to the total count.  

Once the forces in Eq. (\ref{eq:F}) and Eq. (\ref{eq:A}) are evaluated, one uses the Euler integrator
\be
z_i^{\pm}(t+\Delta t) = z_i^{\pm}(t) + \bigg[ v_0 \cos\theta_i(t) - \frac{F(z_i^{\pm}(t))}{\zeta}\bigg] \Delta t,
\ee
plus the expressions for $n_z$ in Eq. (\ref{eq:nz}).


The equations written above actually describe a system of charged parallel sheets moving along 
the $z$-axis and obeying active dynamics in that direction.  The surface charge of each plate is 
$\pm q/A$, where $A=q n/\sigma_c$ is the surface area, and the magnitude of the interaction force 
between any two sheets is $f_{12}=q^2/(2\epsilon A)=q\sigma_c/(2\epsilon n)$.  The strength of the 
interactions, therefore, depends on the number of particles $n$.  (In addition, there is a fixed plate 
at $z=0$ with the surface charge $\sigma_c$).  The mean-field solution  
corresponds to the limit 
$n\to\infty$, where interactions between individual sheets, $f_{12}$, vanish.  If this limit is not satisfied, 
the distributions of the simulated model should no longer correspond to the mean-field approximation.  
To ensure that the procedure and the simulation technique 
is accurate, in Appendix (\ref{sec:anal}) we apply it to the case of passive ions, for which 
analytical expressions are available, so that accuracy can be easily checked.  
The results indicate that already $n=100$ particles
yield accurate mean-field distributions.

The mean-field simulations have previously been used in a number of systems.  
For example, to study the Hamiltonian plasma model comprised of 
Coulomb sheets in a fixed neutralizing background \cite{Dawson62}.  In several versions of the model,
the sheets can either pass through one another or undergo elastic collisions.  A similar technique 
has been applied to study the $XY$-Heisenberg model in periodic boundary conditions 
\cite{Antoni95}.  The method is referred to as the Hamiltonian mean-field model due to the 
underlying Hamiltonian dynamics.  
The Hamiltonian treatment of our system, that is, the switch from active-overdamped to passive-Newtonian dynamics,
would lead to an altogether different physical interpretation, with the latter case representing 
a plasma system.  Such a model should give rise to a system that do not relax to a Boltzmann 
distribution, but for different physical reasons connected with the long-range
interactions and the absence of thermodynamic limit \cite{Yan14}.  

\subsection{calculation of forces and the use of a position index}

Computationally, the most demanding step in the algorithm is the calculation of forces acting on every 
particle at each time step.  
To reduce its computational effort, we implement the following procedure.  
Each particle is assigned two indices, $l$ and $k$.  The first index is simply a label.  
The numbers $1\le l\le n_s+n$ are reserved for counterions 
and $n_s+n < l \le n_T$ for coions.  The other index determines the location of a particle within a sequence, 
\be
0\le z_{n_T} < z_{n_T-1} < \dots < z_k < \dots < z_2 < z_1 < \infty, 
\ee
This index is not permanent but is updated at each time step.  As only a handful of particles exchange their 
relative positions in the sequence, the update is quick and scales as $O(n_T)$.  The procedure consists of a 
loop that compares the positions $z_k$ and $z_{k+1}$ for every $k$.    
If $z_k\ge z_{k+1}$ is false, the two indices are exchanged.  The loop is repeated until the condition
$z_k\ge z_{k+1}$ is true for every $k$.   Once the sequence is established, it is trivial to calculate forces.  
The computation effort of the algorithm scales as $O(n_T)$.

\subsection{reduced units}
In the rest of the paper we use the reduced units.  The dimensionless strength of electrostatic interactions is
\be
\gamma_q = \frac{1}{\zeta v_0} \bigg(\frac{q\sigma_c}{\epsilon}\bigg),
\label{eq:gammaq}
\ee 
and the reduced force is $F^*=F/(\zeta v_0)$.  
Densities are expressed in units $\sigma_c/(ql_p)$, so that the reduced density is 
\be
\rho_{\pm}^* = \rho_{\pm}\bigg(\frac{ql_p}{\sigma_c}\bigg), 
\ee
and the dimensionless salt strength is 
\be
\gamma_{c_s} = c_s\bigg(\frac{ql_p}{\sigma_c}\bigg),
\ee
where $c_s$ is the salt concentration in a bulk.  One can also define a reduced charge density,
\be
\rho_{c}^* = \rho_{c}\bigg(\frac{l_p}{\sigma_c}\bigg).  
\ee

The units of length and time, as before, are the persistence length and time, $l_p$ and $\tau_p$, 
respectively.  
Finally, the Euler integrators become
\be
z_i^{* \pm}(t+\Delta t) = z_i^{* \pm}(t) + \Big[ n_{z_i}(t^*) - \gamma_q A(z_i^{\pm}(t)) \Big] \Delta t^*,
\ee
with $A(z_i^{\pm})$ defined in Eq. (\ref{eq:A}).  Then $n_{z_i}$ is updated according to Eq. (\ref{eq:nz}).

\section{Active counterions}
\label{sec:ocp}

For the counterion-only case the concentration of salt is set to zero, that is, $n_s=0$, then $n_-=n$ and 
$n_+=0$.  Due to dynamic adsorption of counterions on a charged wall, a surface charge 
of a wall becomes reduced, with 
the effective surface charge given as $\sigma_{\rm eff}=g\sigma_c$, where $g=1-n_A/n$ is the 
renormalization constant.  Due to this renormalization, the force exerted on adsorbed particles needs
to be renormalized at each time step, $F(0) = -\hat g\gamma_q$, where $\hat g$ is the instantaneous 
renormalization, such that $g=\langle\hat g\rangle$.  Without adsorption, this force would simply be 
$F(0)=-\gamma_q$.

\begin{figure}[h] 
 \begin{center}
 \begin{tabular}{rr}
  \includegraphics[height=0.19\textwidth,width=0.22\textwidth]{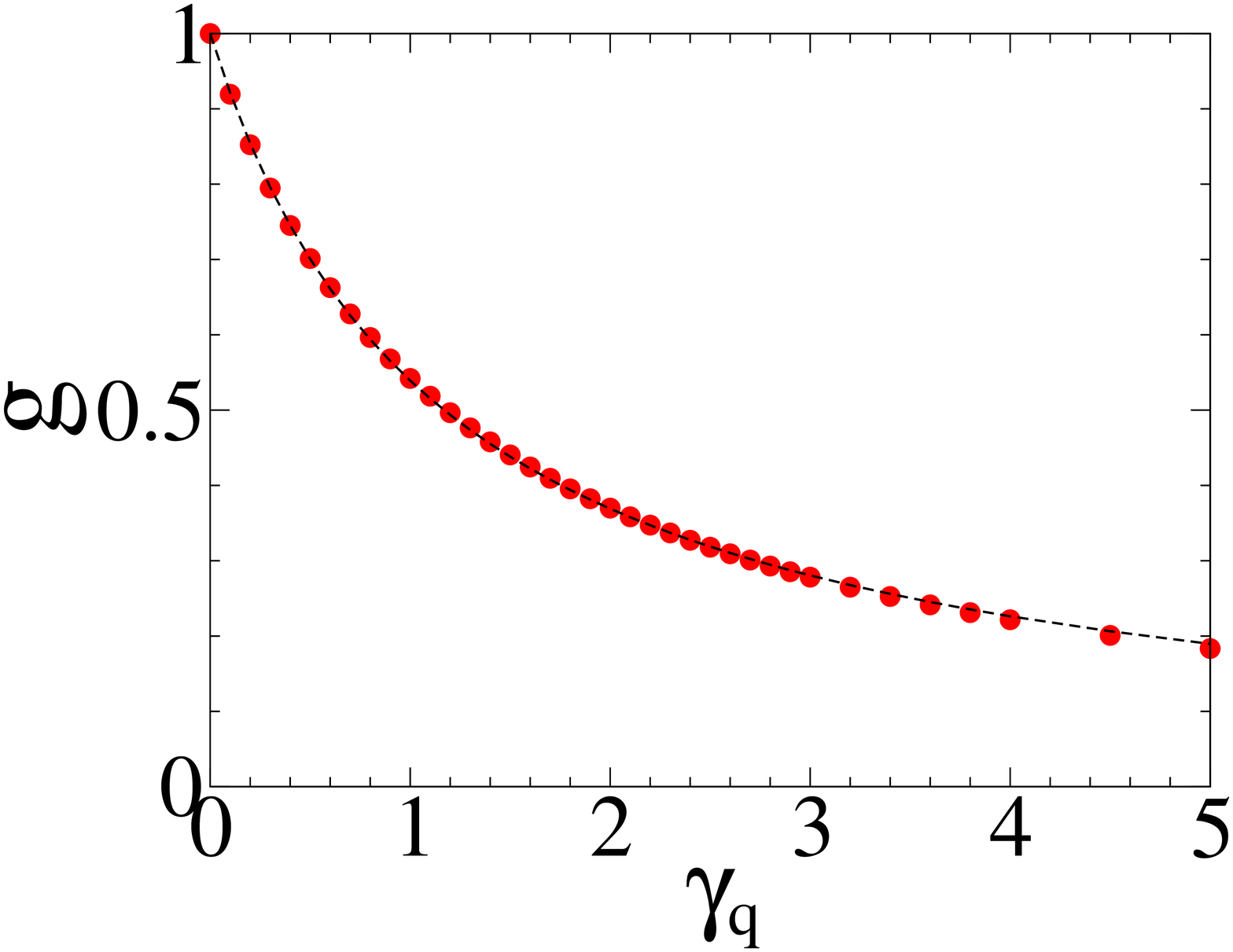}
 \end{tabular}
 \end{center}
\caption{The renormalization constant, $g=1-n_A/n$, as a function of $\gamma_q$.  The data points obtained 
from the mean-field simulation are fitted to a functional form $g=(1+\gamma_q/\gamma_{\rm sat})^{-1}$, where 
$\gamma_{\rm sat}\approx 1.17$.} 
\label{fig:g}
\end{figure}

In Fig. (\ref{fig:g}) we plot $g$ as a function of $\gamma_q$.  The plot indicates that the effective surface 
charge decreases with the strength of electrostatic interactions, and a simple fit to the data points reveals 
the decay rate of $g$ to be algebraic, as $\sim\gamma_q^{-1}$.  Consequently, the effective surface charge 
\graphicspath{{figures/}}
$\sigma_{\rm eff}$ must saturate in the limit $\sigma_c\to\infty$ (see the definition for $\gamma_q$ in 
Eq. (\ref{eq:gammaq})).

Before analyzing the distribution of active counterions, we review 
the mean-field results for passive ions.  The mean-field charge distribution of passive counterions is
\be
\rho_c(z) = \frac{\sigma_c/\lambda_{GC}}{(1 + z/\lambda_{GC})^2},
\label{eq:rho_pb}
\ee  
where 
\be
\lambda_{\rm GC} = \frac{2\epsilon k_BT}{e^2\sigma_c}
\label{eq:lambdaGC}
\ee  
is the Gouy-Chapman length.  Using Eq. (\ref{eq:kT}) to substitute for $k_BT$, 
the Gouy-Chapman length for active particles is 
\be
\frac{\lambda_{\rm GC}}{l_p} = \frac{2}{3}\frac{1}{\gamma_q}. 
\label{eq:lambdaGC2}
\ee

\begin{figure}[h] 
 \begin{center}
 \begin{tabular}{rrr}
  \includegraphics[height=0.19\textwidth,width=0.23\textwidth]{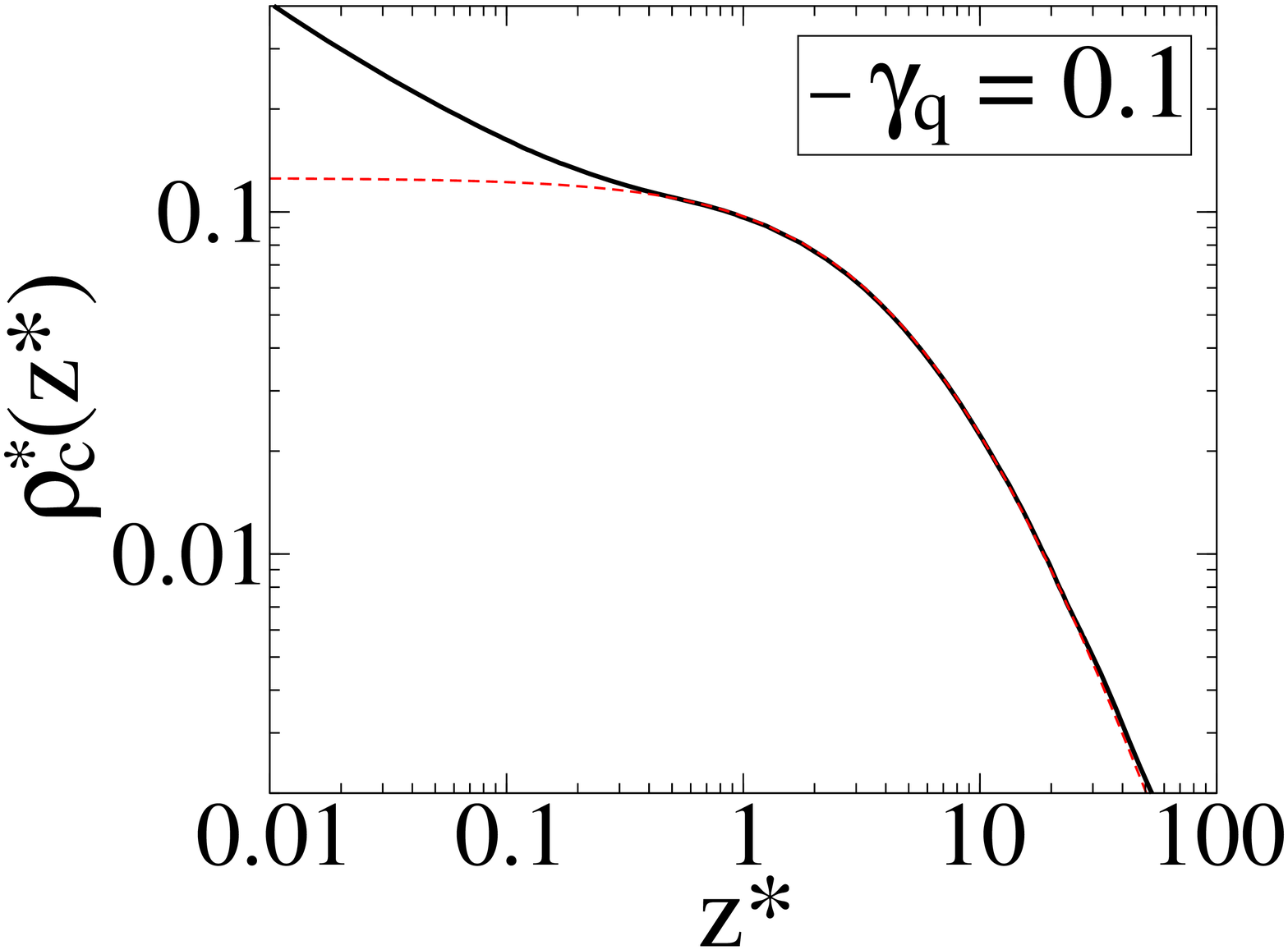}&&
 \end{tabular}
 \end{center}
\caption{Density distribution of counterions for $\gamma_q=0.1$.  The profile obtained from 
a simulation is compared to a fit in Eq. (\ref{eq:rho_fit}) with $g_{\rm fit}\approx 0.91$, determined 
by matching of the far-field regions.  }
\label{fig:rhoA1}
\end{figure}

In the limit $\gamma_q\to 0$ the distributions for passive and active particles converge, apart for the 
divergence at $z=0$ which is absent for passive ions.  In Fig. (\ref{fig:rhoA1}) we plot $\rho_c(z)$, obtained 
from a simulation for $\gamma_q=0.1$, and compare it with the ansantz 
\be
\rho_c(z) = \frac{g^2_{\rm fit} \sigma_c/\lambda_{GC}}{(1+g_{\rm fit} z/\lambda_{GC} )^2},
\label{eq:rho_fit}
\ee
where the fitting parameter for $\gamma_q=0.1$ is $g_{\rm fit}\approx 0.91$.  A small renormalization  
is the result of ion adsorption.  
\graphicspath{{figures/}}

\begin{figure}[h] 
 \begin{center}
 \begin{tabular}{rrr}
  \includegraphics[height=0.19\textwidth,width=0.22\textwidth]{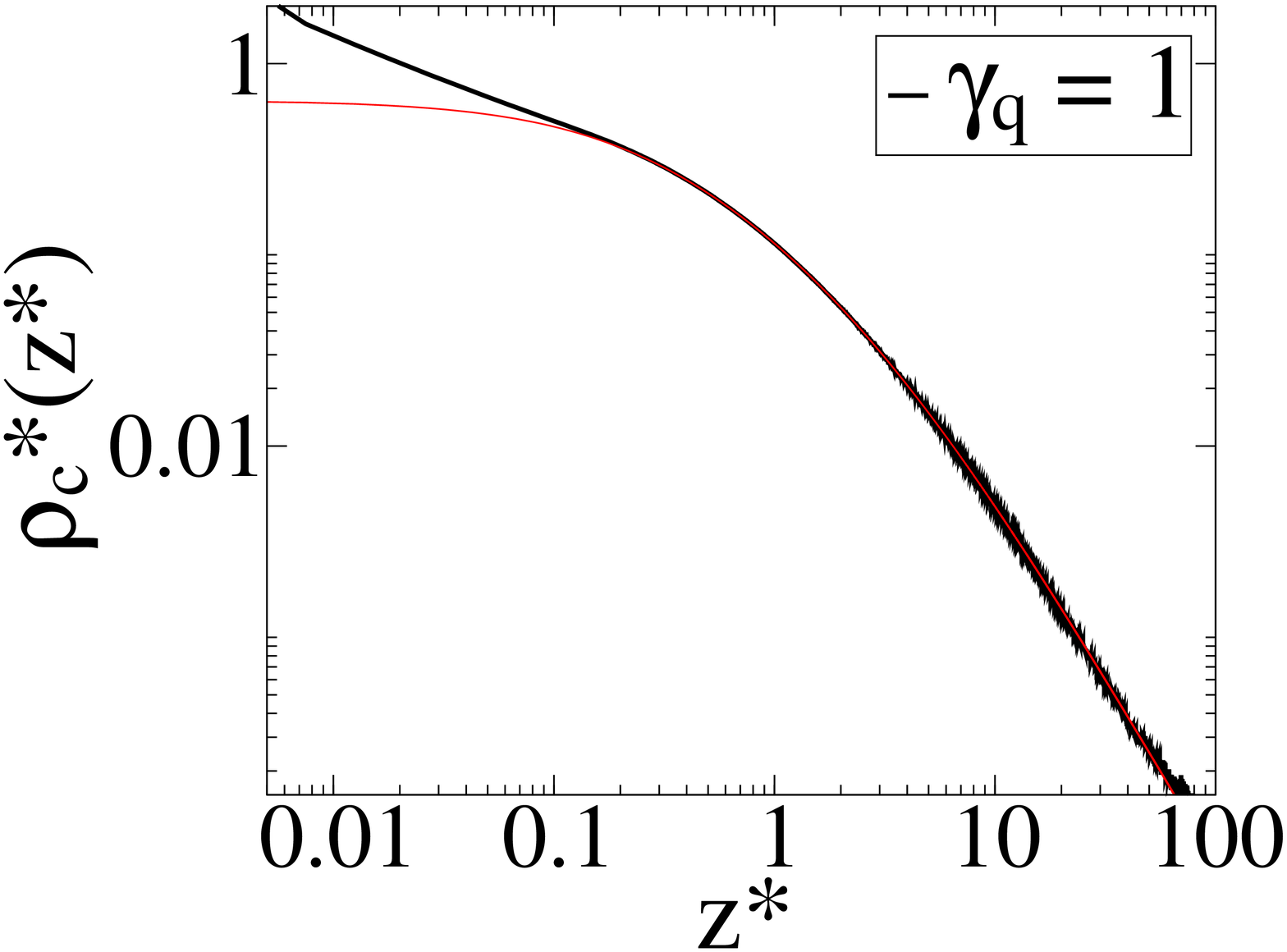}\\
 \end{tabular}
 \end{center}
\caption{Density distribution of counterions for $\gamma_q=1$ together with a fit $\rho_c^*(z) = \frac{a}{1+ bz^* + cz^{*2}}$. } 
\label{fig:rhoA2}
\end{figure}

The ansatz in Eq. (\ref{eq:rho_fit}) becomes less accurate for larger $\gamma_q$.  
A more accurate fit is $\rho_c^*(z) = \frac{a}{b+ cz^* + z^{*2}}$, see Fig. (\ref{fig:rhoA2}).  The ansatz does not 
modify the asymptotic behavior, which is still $\sim z^{-2}$, but it is significantly more accurate in the intermediate 
region.
\graphicspath{{figures/}}
Eq. (\ref{eq:lambdaGC2}), obtained from the mean-field for passive ions, suggests $a=2/(3\gamma_q)$.  
We find this relation to be correct not only in the limit $\gamma_q\to 0$, but for any arbitrary value of $\gamma_q$.  
This then means that the asymptotic region is not modified by counterion adsorption, and far away from a 
wall the distinction between active vs. passive dynamics becomes irrelevant -- at least for a 
counterion-only system.

\section{Active electrolyte}
\label{sec:electrolyte}

Next we consider a charged wall in contact with electrolyte.  
The concentration of salt $c_s$, in the mean-field simulation, is controlled through $n_s$. 
Dissociation of salt brings into the system additional counterions but also coions.  
Because a system is filled with salt ions, one must account for tbe osmotic pressure and its contribution
to dynamic adsorption of counterions as well as the concurrent adsorption of coions.

\begin{figure}[h] 
 \begin{center}
 \begin{tabular}{rr}
  \includegraphics[height=0.19\textwidth,width=0.22\textwidth]{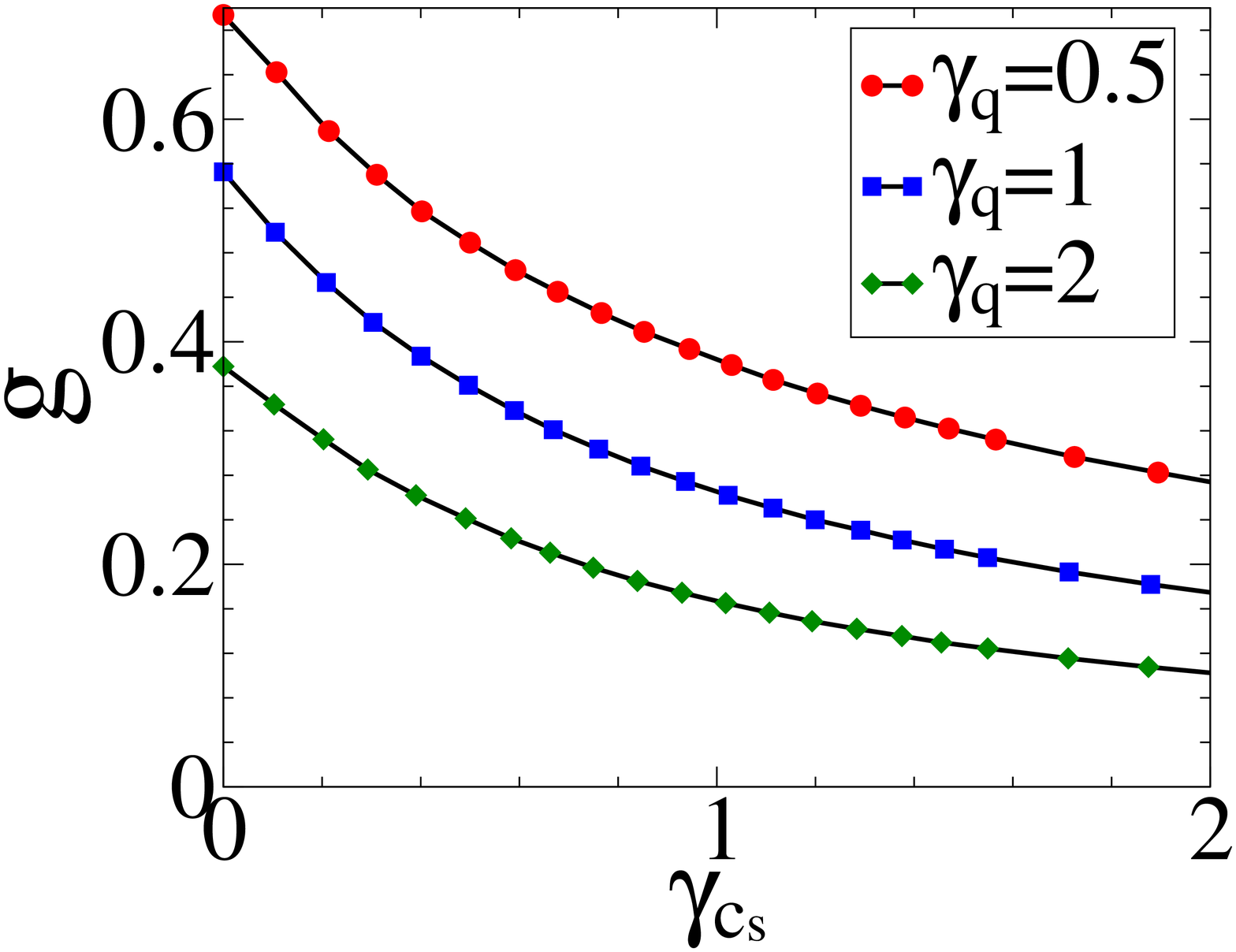}
 \end{tabular}
 \end{center}
\caption{The renormalization constant, $g=1-(n_A^--n_A^+)/n$, as a function of $\gamma_{c_s}$ for 
different values of $\gamma_q$.  The results are from the mean-field simulation.  } 
\label{fig:g2}
\end{figure}

The renormalization factor now is defined as $g = 1-(n_A^--n_A^+)/n$, 
where $n_A^-$ and $n_A^+$ is the number of adsorbed counterions and coions, respectively. 
Fig. (\ref{fig:g2}) shows simulation data points for $g$ as a function of $\gamma_{c_s}$ for a number of 
different $\gamma_{c_s}$.  
The results indicate increased adsorption of charge with increased concentration of salt: the larger 
the concentration of salt, the larger the osmotic pressure pushing particles toward a wall.  Yet the suppression 
of charge is slow due to the concurrent adsorption of coions which reverse the effect of adsorbed counterions.  
\graphicspath{{figures/}}

\begin{figure}[h] 
 \begin{center}
 \begin{tabular}{rr}
  \includegraphics[height=0.19\textwidth,width=0.22\textwidth]{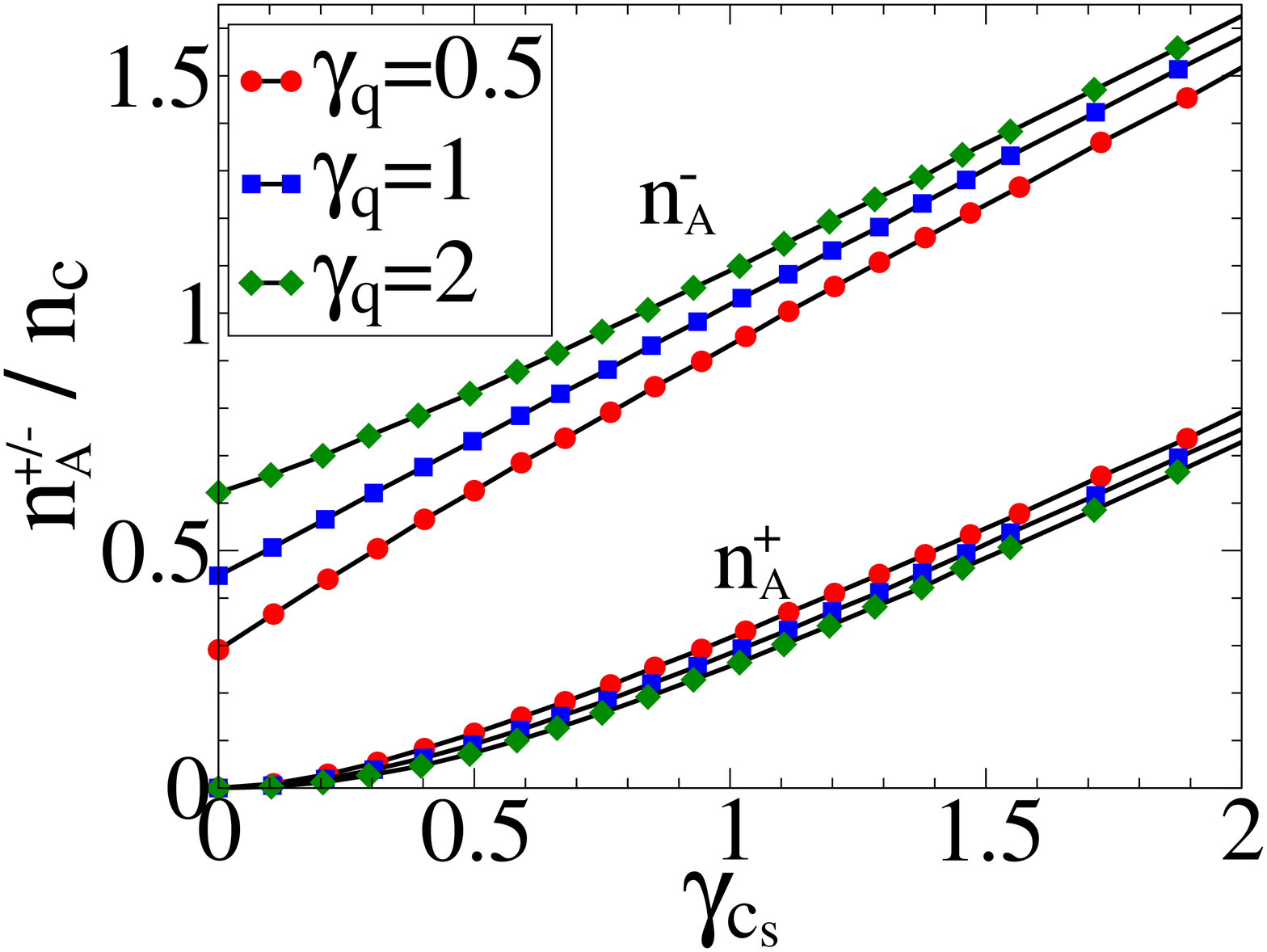}
 \end{tabular}
 \end{center}
\caption{The adsorption of counterions, $n_A^-$, and coions, $n_A^+$, as a function of $\gamma_{c_s}$ for 
different $\gamma_q$. } 
\label{fig:nA}
\end{figure}

In Fig. (\ref{fig:nA}) we plot separate contributions of $g$, that is, $n_A^-$ and $n_A^+$, as a function 
of $\gamma_{c_s}$.  There are a number of interesting observations.  First, the number of adsorbed counterions 
exceeds that of a surface charge, that is $n_A^-/n>1$, at roughy $\gamma_{c_s}>1$, which by itself should 
imply charge inversion.  However, charge inversion is prevented by the concurrent adsorption of coions.  
\graphicspath{{figures/}}
If we look at the adsorption rates, that is, $r^-=\frac{d}{d\gamma_{c_s}}\big(\frac{n_A^-}{n}\big)$ and 
$r^+=\frac{d}{d\gamma_{c_s}}\big(\frac{n_A^+}{n}\big)$ , we find that $r^-$ approaches $r$ (which is the rate of 
adsorption for neutral particles determined to be $r\approx 0.55$) from above, while $r^+$ approaches $r$ from 
below.  In the limit $\gamma_{c_s}\to\infty$ the two rates converge, $r^-=r^+=r$, implying that adsorption is 
entirely determined by osmotic pressure.


Next we look into the far-field region of charge distributions.  
For passive ions, the Debye screening parameter $\kappa$ of the exponential decay 
(related to a screening length as $\lambda_D=\kappa^{-1}$) is 
\be
\kappa = \sqrt{\frac{2c_s q^2}{\epsilon k_BT}}
\ee
Using Eq. (\ref{eq:kT}) to substitute for $k_BT$, we get 
\be
l_p \kappa 
=  \sqrt{6\gamma_q \gamma_{c_s}}.  
\label{eq:kappa}
\ee

For the counterion-only case, the far-field algebraic decay agreed with that for passive 
ions, so that the profiles for passive and active ions differ only in the near-field and intermediate regions.   
Whether Eq. (\ref{eq:kappa}) accurately describes the far-field decay for electrolytes, we examine next.

\begin{figure}[h]
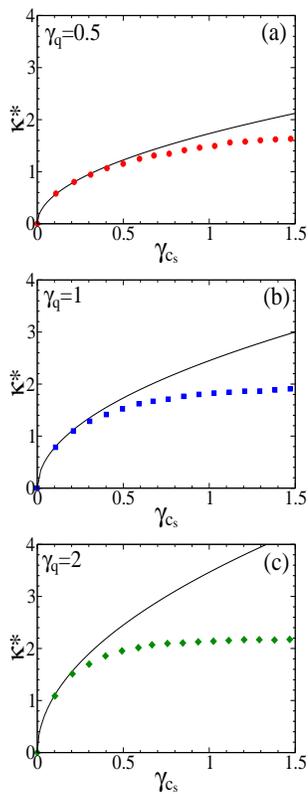
 
 \begin{center}
 \begin{tabular}{rr}
  \includegraphics[height=0.19\textwidth,width=0.22\textwidth]{kappa05.eps}\\
  \includegraphics[height=0.19\textwidth,width=0.22\textwidth]{kappa10.eps}\\
  \includegraphics[height=0.19\textwidth,width=0.22\textwidth]{kappa20.eps}
 \end{tabular}
 \end{center}
\caption{ The screening parameter $\kappa$ for active ions.  The data points from simulations are plotted 
against the analytical expression in Eq. (\ref{eq:kappa}).} 
\label{fig:kappa}
\end{figure}

In Fig. (\ref{fig:kappa}) we plot $\kappa$ obtained by fitting charge distributions to the 
functional form $\rho(z)\sim e^{-\kappa z}$, and then compare the plots 
with the expression in Eq. (\ref{eq:kappa}).  For low concentrations of salt, $\gamma_{c_s}\to 0$, 
the results for active particles agree with Eq. (\ref{eq:kappa}).  
But as $\gamma_{c_s}$ increases, the two plots 
begin to deviate, where active ions show reduced screening, which 
implies a more diffuse double-layer.  The screening parameter, furthermore, appears to 
saturate at a value slightly above $\kappa^* =  2$.  
\graphicspath{{figures/}}

A plausible explanation of a more diffuse double-layer in an active system is some sort of competition between the 
screening and the persistence length.  A persistence length sets a limit on how compressed a double-layer may become.

As the last point, we provide the mean-field expression of pressure for electrolytes,  
\be
\bigg(\frac{q}{\zeta v_0\sigma_c}\bigg) P = \frac{1}{2}\gamma_q + \frac{2}{3}\gamma_{c_s}, 
\ee
where the first term accounts for electrostatic contributions and the second term accounts for 
an (ideal-gas) osmotic pressure.

\subsection{passive coions}

In the above example all ions are active.  In this section we consider a mixture of passive coions and 
active counerions.  As passive coions do not get adsorbed, they don't contribute to the renormalization
of a surface charge.  We set the diffusion constant of passive coions to be the same as that of active 
counterions, $D=D_{\rm eff}$, with $D_{\rm eff}$ defined in Eq. (\ref{eq:Deff}).

\begin{figure}[h] 
 \begin{center}
 \begin{tabular}{rr}
  \includegraphics[height=0.19\textwidth,width=0.22\textwidth]{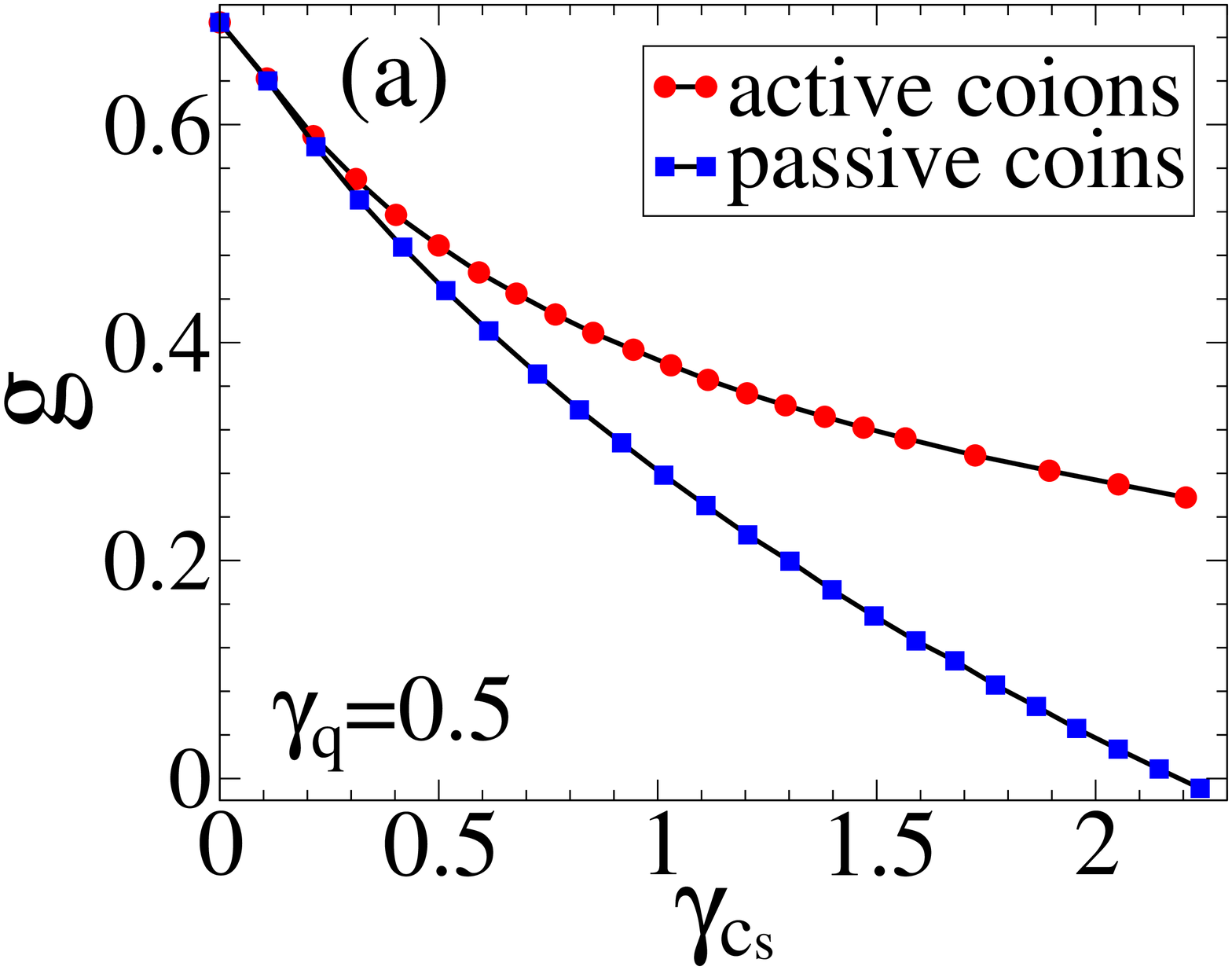}
  \includegraphics[height=0.19\textwidth,width=0.22\textwidth]{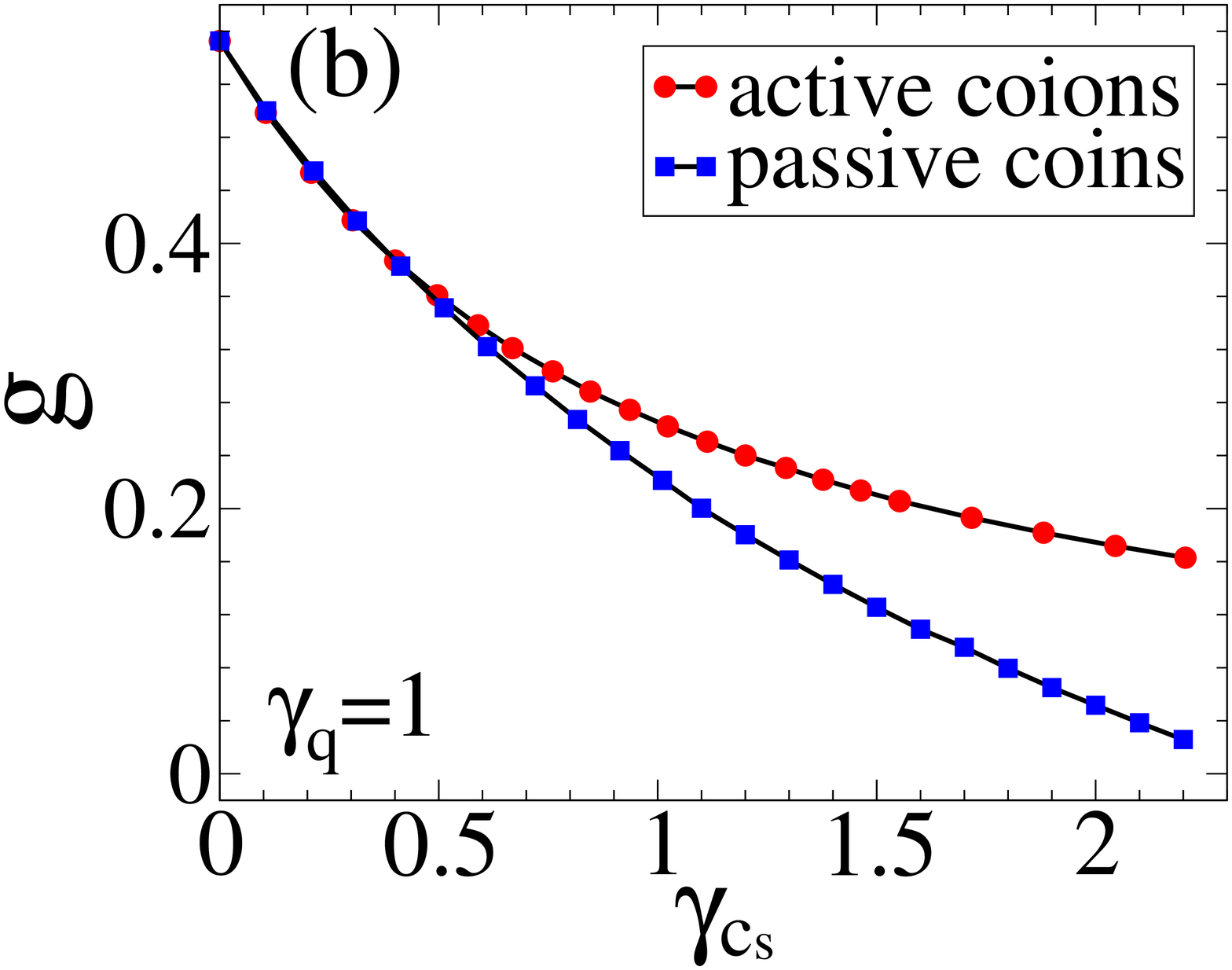}
 \end{tabular}
 \end{center}
\caption{Renormalization constant $g$ as a function of $\gamma_{c_s}$ for $\gamma_q=0.5$ 
and $\gamma_q=1$.  } 
\label{fig:g3}
\end{figure}

In Fig. (\ref{fig:g3}) we plot the renormalization constant, $g=1-n_A^-/n$, for the 
above described system.  
\graphicspath{{figures/}}
In comparison to the results in Fig. (\ref{fig:g2}), the renormalization constant becomes zero at 
$\gamma_{c_s}\approx 2.2$ (for $\gamma_q=0.5$) and $\gamma_{c_s}\approx 2.4$ (for $\gamma_q=1$), then
beyond, $g$ changes sign, indicating a "surface charge inversion" as adsorbed 
counterions overcompensate a bare surface charge.

\begin{figure}[h] 
 \begin{center}
 \begin{tabular}{rr}
  \includegraphics[height=0.19\textwidth,width=0.22\textwidth]{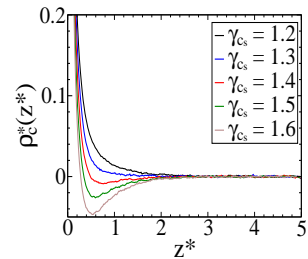}
 \end{tabular}
 \end{center}
\caption{The charge density profiles for $\gamma_q=0.5$, for different values of $\gamma_{c_s}$.  
There is an indication of counterion overscreening.  } 
\label{fig:rhocd}
\end{figure}

In Fig. (\ref{fig:rhocd}) we plot charge densities for different values of $\gamma_{c_s}$ (for $\gamma_q=0.5$).  
At $\gamma_{c_s} = 1.3$ the charge distribution changes sign, even before the "surface charge inversion"
at $\gamma_{c_s}\approx 2.2$.  We refer to this as "overscreening", to distinguish it from "surface charge inversion".  
"Overscreening" can be traced to the divergent density profile, whose emergence is concurrent with dynamic adsorption.  
Even if a surface charge has not yet changed a 
sign, the newly released counterions that linger in the vicinity of a wall, together with adsorbed counterions, 
overcompensate a bare surface charge.  "Overscreening", therefore, precedes and anticipates "surface charge inversion".  
\graphicspath{{figures/}}

\subsection{passive counterions}

\begin{figure}[h] 
 \begin{center}
 \begin{tabular}{rr}
  \includegraphics[height=0.19\textwidth,width=0.22\textwidth]{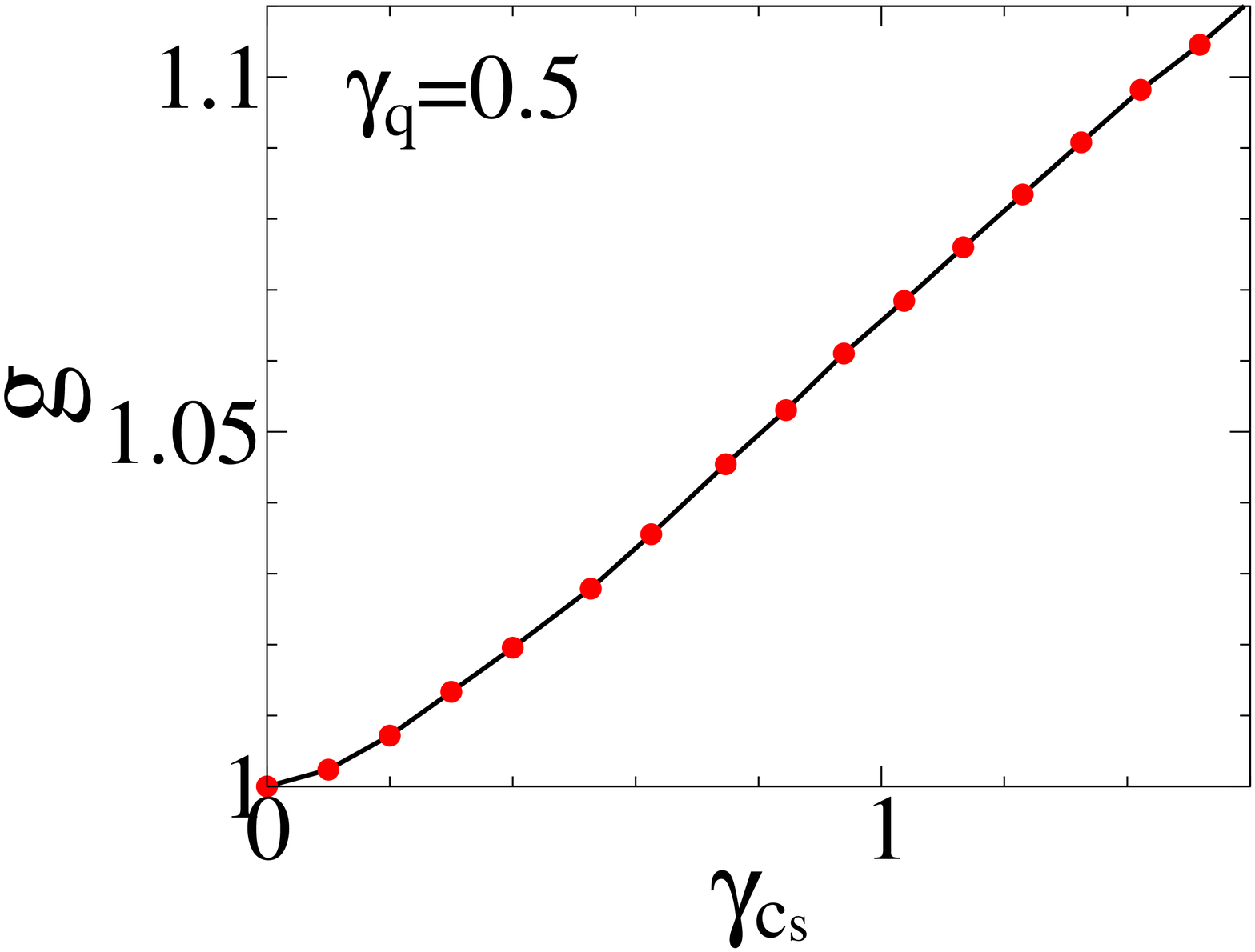}
 \end{tabular}
 \end{center}
\caption{Renormalization constant $g$ as a function of $\gamma_{c_s}$ for an electrolyte mixture 
of active coions and passive counterions.  } 
\label{fig:g4}
\end{figure}

As the last case study, we consider a mixture of active coions and passive counterions.  In this case only 
coions become adsorbed, and the adsorbed coions renormalize a surface charge up.  
The renormalization constant for this situation is defined as $g=1+n_A^+/n$.  
Fig. (\ref{fig:g4}) shows $g$ as a function of $\gamma_{c_s}$.  As coions are generally depleted 
from a charged surface, the coion adsorption and the resulting charge renormalization
is considerably weaker than that caused by active counterions.  
\graphicspath{{figures/}}


\section{Conclusion}
\label{sec:conclusion}

In the present work we investigate a new type of adsorption that arises in active
dynamics and does not depend on the chemistry of a surface, as is the case with 
chemisorption and physiosorption.  The dynamic adsorption 
is particularly relevant for electrolytes, where the counterion adsorption 
renormalizes a surface charge down, and the coion adsorption renormalizes it up.  
The counterion adsorption, under some
conditions, may lead to surface charge inversion.  Then a divergent density near a 
charged wall, which is concurrent with dynamic adsorption, can lead to overscreening of
a surface charge, prior to surface charge inversion.  
Far away from a wall, active dynamics modifies the screening of a surface charge, 
leading to a more diffuse double-layer.

While these are all strictly dynamic 
effects, without direct counterpart in the passive system, there are some similarities with the
behavior of Coulomb fluids in the vicinity of charge-disordered surfaces (diverging surface 
density of mobile charge \cite{Rudi15a}).

Based on the above conclusions one can state that the properties of active Coulomb fluids are 
unexpected and warrant further consideration, especially since addition of active charged 
particles to a colloid solution could pave the way to control and modify the electrostatics of 
colloids in the way that was impossible to contemplate from the mere passive particle perspective.

\begin{acknowledgments}
D.F. would like to acknowledge discussions with Marco Leoni on the subject of 
active charges, and the remote use of computational 
resources in the ESPCI via a generous permission of Tony Maggs and Michael Schindler.  
R.P. was supported by the 1000-Talents Program of the Chinese Foreign Experts Bureau.
\end{acknowledgments}

\appendix
\section{The Fokker-Planck equation for active particles}
\label{sec:anal0}
The probabilty distribution $\rho({\bf r},\theta,\phi,t)$ of non-interactive active particles satisfy the following 
Fokker-Planck equation
\ba
\frac{\partial\rho}{\partial t} 
&=& -\bnabla\cdot\left[\left(v_0 {\bf n} (\theta,\phi) 
+ \frac{{\bf F}({\bf r})}{\zeta}\right)\rho\right]  \nonumber\\
&+&\frac{D_r}{\sin\theta}\frac{\partial}{\partial\theta}\bigg[\sin\theta\frac{\partial\rho}{\partial\theta}\bigg]
+ \frac{D_r}{\sin^2\theta}\frac{\partial^2\rho}{\partial\phi^2},
\label{eq:FP}
\ea
where ${\bf F}({\bf r})$ is the external force.  The above equation comprises two processes.  
The first process is simple convection, $-\bnabla\cdot {\bf j}$, with the flux given by 
\be
{\bf j} ({\bf r},\theta,\phi,t) = \left(v_0 {\bf n}(\theta,\phi) + \frac{{\bf F}({\bf r})}{\zeta}\right)\rho({\bf r},\theta,\phi,t),  
\ee
where the flow is due to intrinsic particle force $\zeta v_0$ and an external force field ${\bf F}({\bf r})$.  
The second process is the diffusion of a director vector ${\bf n}$ represented by an angular part of the Laplace 
operator with radius set to one.  The combination of the two processes gives rise to active transport.  

For a wall geometry considered in this work, where particles are confined to a half-space $z\ge 0$ and 
where an external force acts only in the $z$-direction, ${\bf F}=(0,0,F)$, the relevant distribution is 
$\rho(z,\theta,t)$.  Furthermore, as we focus on stationary distributions, the relevant differential equation is 
\ba
&&\frac{\partial}{\partial z}\left[\left(v_0\cos\theta  + \frac{F(z)}{\zeta}\right)\rho(\theta,z)\right]  = \nonumber\\ 
&&~~~~~~~~~~~~~~~~~~~~~~~~~~~~~~
\frac{D_r}{\sin\theta}\frac{\partial}{\partial\theta}\bigg[\sin\theta\frac{\partial\rho(\theta,z)}{\partial\theta}\bigg].
\label{eq:FP1Da}
\ea

The application of the method of separation of variables assumes $\rho(z^*,\theta)=\Gamma(z^*)T(\theta)$.  
For non-interactive active particles in gravitational field the separation of variables yields two equations, 
\be
\frac{\partial\Gamma_n(z^*)}{\partial z^*}  + 3\gamma_G\lambda_n \Gamma_n(z^*) = 0, 
\label{eq:SV1}
\ee
\be
\frac{\partial}{\partial\theta}\bigg[\sin\theta\frac{\partial K_n(\theta)}{\partial\theta}\bigg]
+ 6\gamma_G\lambda_n \sin\theta(\cos\theta - \alpha) K_n(\theta) = 0.  
\label{eq:SV2}
\ee

\section{The mean-field simulation of passive ions: 
analytical results}
\label{sec:anal}

In this section we apply the mean-field simulation to passive counterions (or counterion sheets).  
For passive Brownian particles the weight of every configuration is proportional to the 
Boltzmann factor $e^{- U(z_1,\dots,z_n)/k_BT}$, where $U(z_1,\dots,z_n)$ is the total electrostatic 
potential of the system.  Analysis becomes more tractable if sheets are arranged into sequence 
\cite{Lenard61,Lenard62},  
\be
0\le z_n\le z_{n-1}\le \dots \le z_2\le  z_1<\infty, 
\ee
so that the index $i$ functions both as label and position index.  
A pair potential between any two sheets $i$ and $j$ is 
\be
\frac{u(z_i,z_j)}{k_BT} = \left\{ 
  \begin{array}{r l}
    (z_i-z_j)/(n\lambda_{GC}), & \quad \text{if $i<j$}\\
     (z_j-z_i)/(n\lambda_{GC}), & \quad \text{if $i>j$},
             \label{eq:uij}
  \end{array} 
  \right.
  \ee
where 
$\lambda_{GC}$ is the Gouy-Chapman length defined in Eq. (\ref{eq:lambdaGC}). The factor $1/n$ 
indicates that the larger the number of sheets, the weaker the pair interactions.  
Taking into account the electrostatic potential due to a fixed charged wall at $z=0$, 
$u_{\rm wall}(z_i)/k_BT = z_i/\lambda_{GC}$, the total potential energy is found to be 
\be
\frac{U(z_1,\dots,z_n)}{k_BT} =  \frac{1}{n} \sum_{k=1}^n(2k-1) \frac{z_k}{\lambda_{GC}},
\label{eq:U}
\ee
where the plate $k=1$ feels the weakest potential, and the plate $k=n$ feels the 
strongest potential.  The partition function can now be written as 
\be
Z_n= 
\prod_{k=1}^{n} \int_0^{z_{k-1}} \!\!\!\!\! dz_k\, e^{-(2k-1) z_k/(n\lambda_{GC})} 
= \frac{n^n\lambda_{GC}^n}{n!n!}
\label{eq:ZN}
\ee
where $z_0=\infty$.   The limits of the integrals prevent positional permutations between sheets, which makes
particles distinguishable, and there is no need of the factor $1/n!$.

Because sheets occupy different positions within a sequence, their distributions are unique, given by 
\ba
&& p_{m}^{(n)}(z_m)  = \nonumber\\
&&~~~~~~\frac{e^{- (2m-1)z_m / (n\lambda_{GC})}  }{Z_n} 
\prod_{k=1}^{m-1} \int_{z_m}^{z_{k-1}} \!\!\!\!\! dz_k\, e^{-(2k-1) z_k/(n\lambda_{GC})} \nonumber\\
&&~~~~~~\times \prod_{k=m+1}^{n} \int_{0}^{z_{k-1}} \!\!\!\!\! dz_k\, e^{-(2k-1) z_k/(n\lambda_{GC})}.  
\nonumber\\
\label{eq:p}
\ea
The term of the first line evaluates to 
\be
\frac{Z_{m-1}}{Z_n} e^{-m^2 z_m/(n\lambda_{GC})},
\ee
and is obtained by shifting all the variables of integration as $x_k = z_k - z_m$, which allows us to write
\ba
&&\prod_{k=1}^{m-1} \int_{z_m}^{z_{k-1}} \!\!\!\!\!\! dz_k\, e^{-(2k-1) z_k/(n\lambda_{GC})}  \nonumber\\
&& ~~~~~~~~~~ =  \prod_{k=1}^{m-1} \int_{0}^{x_{k-1}} \!\!\!\!\!\! dx_k\, e^{-(2k-1) (x_k-z_m)/(n\lambda_{GC})}.  
\nonumber\\
\ea
The term of the second line is more difficult to evaluate but, after some manipulation, we get 
\ba
&&2(-1)^{m}
(n\lambda_{GC})^{n-m} \nonumber\\
&&\times \sum_{k=m}^n \! \frac{(-1)^{k} k e^{-(k^2-m^2)z_m/(n\lambda_{GC})} (k+m-1)!}{(k-m)!(n+k)!(n-k)!}.  
\nonumber\\
\ea
A distribution for a plate $m$ then becomes 
\ba
p_m^{(n)}(z) &=& \frac{1}{n\lambda_{GC}}\frac{2 (-1)^{m} \, n!n!  } {(m-1)!(m-1)!} \nonumber\\
&\times& \sum_{k=m}^n \frac{(-1)^{k} k e^{-k^2 z/(n\lambda_{GC})} (k+m-1)!}{(k-m)!(n+k)!(n-k)!}, 
\nonumber\\
\label{eq:p}
\ea
where we drop the subscript $m$ from $z$. From the identity 
\be
\sum_{k=m}^n \frac{(-1)^{k} (k+m-1)!}{k(k-m)!(n+k)!(n-k)!} = \frac{(-1)^m(m-1)!(m-1)!}{2\, n!n!}, 
\ee
we know that all the distributions $p_m(z)$ are normalized, 
\be
\int_0^{\infty} dz\, p_m^{(n)}(z) = 1. 
\ee

The charge distribution is obtained by summing up the distributions of all the sheets, 
\be
\rho_c^{(n)}(z) = \frac{\sigma_c}{n} \sum_{m=1}^n p_m(z), 
\label{eq:rho_n}
\ee
where the factor $\sigma_c/n$ ensures that $\int_0^{\infty}dz\,\rho_c^{(n)}(z)=\sigma_c$, so that 
the charge density is independent of $n$.
For the case $n=1$, the distribution is exponential, 
\be
\rho_c^{(1)}(z) = \frac{\sigma_c}{\lambda_{GC}} e^{-z/\lambda_{GC}},
\ee
and corresponds to a one-particle distribution that is exactly the counterion density in the strong-coupling 
approximation \cite{Rudi13}.  Then, in the limit $n\to\infty$, we find 
\be
\rho_c^{(n)}(z) = \frac{\sigma_c/\lambda_{GC}}{(1+z/\lambda_{GC})^2} + O\bigg(\frac{1}{n}\bigg),
\ee
where the above expression is obtained by expanding the exponential term $e^{-k^2 z/(n\lambda_{GC})}$ 
in Eq. (\ref{eq:p}).  The dominant term corresponds to the solution of the Poisson-Boltzmann equation 
for the present problem, see Eq. (\ref{eq:rho_pb}).

\begin{figure}[h] 
 \begin{center}
 \begin{tabular}{rr}
  \includegraphics[height=0.21\textwidth,width=0.27\textwidth]{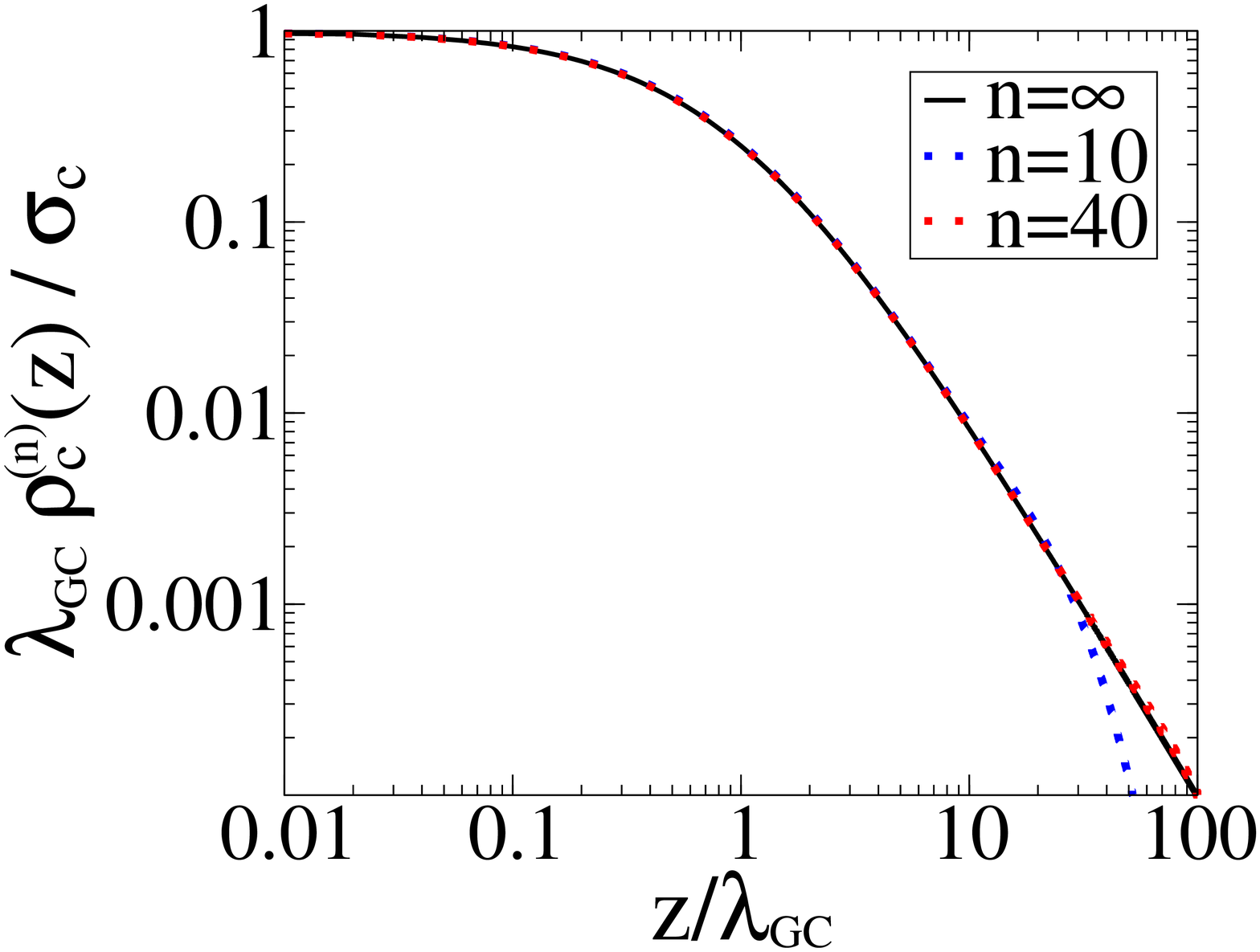}
 \end{tabular}
 \end{center}
\caption{Charge density distributions, $\rho_c^{(n)}(z)$, given in Eq. (\ref{eq:rho_n}), for different 
values of $n$.  } 
\label{fig:rho_n}
\end{figure}

In Fig. (\ref{fig:rho_n}) we plot a number of charge density distributions for different values of $n$.  Already for 
$n=10$ the distribution is accurate up to the point $z/\lambda_{GC}\approx 10$, where the exponential
decay takes over.  For $n=40$ the range of accuracy increases to $z/\lambda_{GC} \approx 100$.  
It appears then that the main size effect is the range of validity.  
For our simulations we use the values between $n=1000$ and 
$n=6000$, depending on the situation.  
\graphicspath{{figures/}}


\begin{thebibliography}{99}

\bibitem{Rudi13}
A. Naji, M. Kandu\^ c, J. Forsman, R. Podgornik, J. Chem. Phys. {\bf 139}, 150901 (2013).

\bibitem{David18}
T. Markovich, D. Andelman and R. Podgornik, in Handbook of Lipid Membranes, ed. C. Safynia and J. Raedler, Taylor \& Francis, 2018.


\bibitem{Yan02}
Y. Levin, Rep. Prog. Phys. {\bf 65}, 1577 (2002). 

\bibitem{Frydel16c}
D. Frydel and M. Ma, Phys. Rev. E {\bf 93}, 062112 (2016).

\bibitem{Frydel17}
Y. Xiang and D. Frydel, J. Chem. Phys. {\bf 146}, 194901 (2017).




\bibitem{David97}
I.~Borukhov, D.~Andelman, H.~Orland,  Phys. Rev. Lett {\bf 79}, 435 (1997).

\bibitem{Frydel11a}
D. Frydel and M. Oettel, Phys. Chem. Chem. Phys. {\bf 13}, 4109 (2011).  

\bibitem{Frydel12}
D. Frydel and Y. Levin, J. Chem. Phys. {\bf 137}, 164703 (2012).  

\bibitem{Frydel14}
D. Frydel and M. Oettel, Contribution to the proceedings of the 6th Liquid Matter Conference (2014).  




\bibitem{David07}
A.~Abrashkin, D.~Andelman, H.~Orland,  Phys. Rev. Lett {\bf 99}, 077801 (2007).

\bibitem{Rudi09}
M. Kandu\v c, A. Naji, Y. S. Jho, P. A. Pincus, R. Podgornik, J. Phys.: Condens Matter. {\bf 21}, 424103 (2009).




\bibitem{Frydel11}
D. Frydel, J. Chem. Phys. {\bf 134}, 0234704 (2011).  

\bibitem{Hatlo12}
M. M. Hatlo, R. van Roij and L. Lue, Eur. Phys. Lett. {\bf 97} 28010 (2012).




\bibitem{Hansen11a}
D. Coslovich, J.-P. Hansen, and G. Kahl, Soft Matter {\bf 7},  1690 (2011).

\bibitem{Hansen11b}
D. Coslovich, J.-P. Hansen, and G. Kahl, J. Chem. Phys. {\bf 134},  244514 (2011).

\bibitem{Frydel13}
D. Frydel and Y. Levin, J. Chem. Phys. {\bf 138}, 174901 (2013).  

\bibitem{Masters13}
P. B. Warren, A. Vlasov, L. Anton, and A. J. Masters, J. Chem. Phys. {\bf 138}, 204907 (2013).  

\bibitem{Levesque14}
J.-M. Caillol and D. Levesque, J. Chem. Phys. {\bf 140},  214505 (2014).  

\bibitem{Frydel16a}
D. Frydel, J. Chem. Phys. {\bf 145}, 184703 (2016).

\bibitem{Frydel18}
D. Frydel and Y. Levin, J. Chem. Phys. {\bf 148}, 024904 (2018).




\bibitem{Bohinc04}
K. Bohinc, A. Igli\v c and S. May, {\sl Europhys. Lett.} {\bf 68}, 494 (2004).

\bibitem{Bohinc08}
S. May, A. Iglic, J. Rescic, S. Maset and K. Bohinc, {\sl J. Phys. Chem. B} {\bf 112}, 1685 (2008).

\bibitem{Bohinc11}
K. Bohinc, J. Rescic, J. Maset and S. May, {\sl J. Chem. Phys. } {\bf 134}, 07411 (2011).

\bibitem{Bohinc12}
K. Bohinc, J. M. A. Grime, L. Lue, {\sl Soft Matter} {\bf 8}, 5679 (2012).





\bibitem{Frydel16b}
D. Frydel,  Adv. Chem. Phys.  {\bf 160}, 209 (2016).



\bibitem{Lee13}
C. F. Lee, N. J. Phys. {\bf 15} (2013).






\bibitem{Ninham71}
Ninham B. W. and Parsegian V. A., J. Theor. Biol., {\bf 31} 405 (1971).

\bibitem{White75}
D. Chan, J. W. Perram, L. R. White and T. W. Healy, J. Chem. Soc. Faraday Trans. I, {\bf 71}, 1046 (1975).



\bibitem{Rudi14}
N. Adzic and R. Podgornik, Euro. Phys. J. E {\bf 37}, 49 (2014).

\bibitem{Rudi15}
N. Adzic and R. Podgornik, Phys. Rev. E, {\bf 91}, 022715 (2015).

\bibitem{Tomer16}
T. Markovich, D. Andelman and R. Podgornik, EPL {\bf 113}, 26004 (2016).

\bibitem{Tomer18}
T. Markovich, D. Andelman and R. Podgornik, EPL {\bf 120}, 26001 (2018).


\bibitem{Blum89}
L. Blum, M. L. Rosinberg, and J. P. Badiali, J. Chem. Phys. {\bf 90}, 1285 (1989).  

\bibitem{McQuarrie95}
J. A. Greathouse and D. A. McQuarrie, J. Colloid Interface Sci. {\bf 175}, 219 (1995).

\bibitem{McQuarrie96a}
J. A. Greathouse and D. A. McQuarrie, J. Colloid Interface Sci. {\bf 181}, 319 (1996).

\bibitem{McQuarrie96b}
J. A. Greathouse, and D. A. McQuarrie,  J. Phys. Chem. {\bf 100}, 1847 (1996).







\bibitem{Fisch80}
N. J. Fisch and M. D. Kruskal, J. Math. Phys. {\bf 21}, 740 (1980).

\bibitem{Cates15c}
A.P.  Solon, M.E. Cates, and J. Tailleur, Eur. Phys. J. Special Topics {\bf 224}, 1231 (2015).  

\bibitem{Wagner17}
C. G. Wagner, M. F. Hagan and A. Baskaran, J. Stat. Mech., 043203 (2017).



\bibitem{Lowen16}
C. Bechinger, R. Di Leonardo, H. L\"owen, C. Reichhardt, G. Volpe, and G. Volpe, Rev. Mod. Phys. {\bf 88}, 045006 (2016).




\bibitem{Brady16}
S. C. Takatori, R. Dier, J. Vermant and J. F. Brady, Nature Communications {\bf 7}, 10694 (2016).



\bibitem{Chen2014}
J. Gao, P. Tang, Y. Yanga and J. Z. Y. Chen, Soft Matter {\bf 10}, 4674 (2014).

\bibitem{Chen2016}
J. Z. Y. Chen, Prog. Pol. Sci. {\bf 54}, 3 (2016).



\bibitem{Hsu03}
E. P. Hsu. A Brief Introduction to Brownian Motion on a Riemannian Manifold. Summer School in Kyushu, (2008).

\bibitem{Engel04}
M. Raible and A. Engel, Appl. Organometal. Chem. {\bf 18}, 536 (2004).

\bibitem{Gompper15}
R. G. Winkler, A. Wysocki and G. Gompper, Soft Matter {\bf 11}, 6680 (2015).



\bibitem{Mathieu1868}
\'E. Mathieu, J. Math. Pure Appl. {\bf 13}, 137 (1868).

\bibitem{Ruby96}
L. Ruby, Am. J. of Phys. {\bf 64}, 39 (1996). 

\bibitem{Frenkel01}
D. Frenkel and R. Portugal, J. Phys. A: Math. Gen. {\bf 34} 3541 (2001).












\bibitem{Brady14}
S. C. Takatori, W. Yan, and J. F. Brady, Phys. Rev. Lett. {\bf 113}, 028103 (2014).

\bibitem{Cates15a}
A. P. Solon, Y. Fily, A. Baskaran, M. E. Cates, Y. Kafri, M. Kardar and J. Tailleur, Nature Physics {\bf 11}, 673 (2015).

\bibitem{Cates15b}
A. P. Solon, J. Stenhammar, R. Wittkowski, M. Kardar, Y. Kafri, M. E. Cates, J. Tailleur, Phys. Rev. Lett. {\bf 114}, 198301 (2015).

\bibitem{Dean18}
M. Kr\"uger, A. Solon, V. D\'emery, C. M. Rohwer, and D. S. Dean, (2018).














\bibitem{Dawson62}
J. Dawson, Phys. Fluids {\bf 5}, 445 (1962).

\bibitem{Antoni95}
M. Antoni, S. Ruffo, Phys. Rev. E {\bf 52}, 2361 (1995).  

\bibitem{Yan14}
Y. Levin, R. Pakter, F. B. Rizzato, T. N.Teles, F. P.C.Benetti, Phys. Rep. {\bf 535}, 1 (2014).







\bibitem{Rudi15a}
M. Ghodrat, A. Naji, H. Komaie-Moghaddam, and R. Podgornik, J. of Chem. Phys. {\bf 143}, 234701 (2015).  








\bibitem{Lenard61}
A. Lenard, J. Math. Phys. {\bf 2}, 682 (1961).

\bibitem{Lenard62}
S.F. Edwards and A. Lenard, J. Math. Phys. {\bf 3}, 778  (1962). 






























\end{thebibliography}
\end{document}